\documentclass[twocolumn]{aastex63} 

\usepackage{graphicx} 
\usepackage{float} 
\usepackage{wrapfig} 
\usepackage{lipsum} 
\usepackage{hyperref}
\usepackage{times}
\usepackage{amsmath}
\usepackage{graphicx}
\usepackage{subfigure}
\usepackage{placeins}
\usepackage{hyperref}
\usepackage{gensymb}
\usepackage{upgreek}
\usepackage{natbib}

\usepackage{graphicx}
\usepackage{subfigure}
\usepackage{multirow}
\usepackage{comment}
\usepackage{natbib}
\usepackage{hyperref}
\usepackage{mathtools}
\usepackage{mathrsfs}
\usepackage{fontenc}
\usepackage{color}
\usepackage{url}
\usepackage{hyperref}
\usepackage{gensymb}
\usepackage{pifont}
\graphicspath{{./}}

\newcommand       \K            {\,{\rm K}}
\newcommand       \mum          {{\rm \mu m}}

\submitjournal{ApJ}

\shorttitle{Shocks Destroy PAHs in Low-luminity AGNs}
\shortauthors{Zhang, Ho, and Li}

\begin{document}

\title{Evidence that Shocks Destroy Small PAH Molecules in Low-luminosity Active Galactic Nuclei}

\author[0000-0003-4937-9077]{Lulu Zhang}
\affiliation{Kavli Institute for Astronomy and Astrophysics, Peking University, Beijing 100871, China; l.l.zhang@pku.edu.cn}
\affiliation{Department of Astronomy, School of Physics, Peking University, Beijing 100871, China}

\author[0000-0001-6947-5846]{Luis C. Ho}
\affiliation{Kavli Institute for Astronomy and Astrophysics, Peking University, Beijing 100871, China; l.l.zhang@pku.edu.cn}
\affiliation{Department of Astronomy, School of Physics, Peking University, Beijing 100871, China}

\author[0000-0002-1119-642X]{Aigen Li}
\affiliation{Department of Physics and Astronomy, University of Missouri, Columbia, MO 65211, USA}

\begin{abstract}

We combined mapping-mode mid-infrared Spitzer spectra with complementary infrared imaging to perform a spatially resolved study of polycyclic aromatic hydrocarbons (PAHs) emission from the central regions of 66 nearby galaxies, roughly evenly divided into star-forming systems and low-luminosity active galactic nuclei (AGNs).  In conjunction with similar measurements available for quasars, we aim to understand the physical properties of PAHs across a broad range of black hole accretion power, with the goal of identifying observational diagnostics that can be used to probe the effect of AGNs on the host galaxy.  Whereas the PAH emission correlates tightly with far-ultraviolet luminosity in star-forming regions, the spatially resolved regions of AGNs tend to be PAH-deficient. Moreover, AGN regions exhibit on average smaller PAH 6.2\,$\mum$/7.7\,$\mum$ and larger PAH 11.3\,$\mum$/7.7\,$\mum$ band ratios. Although the current data are highly restrictive, they suggest that these anomalous PAH band ratios cannot be explained by the effects of the AGN radiation field alone. Instead, they hint that small grains may be destroyed by the combined effects of radiative processes and shocks, which are plausibly linked to jets and outflows preferentially associated with highly sub-Eddington, radiatively inefficient AGNs.  While quasars also present a PAH deficit and unusual PAH band ratios, their characteristics differ in detail compared to those observed in more weakly accreting AGNs, a possible indicator of fundamental differences in their modes of energy feedback.
\end{abstract}


\section{Introduction} 

Polycyclic aromatic hydrocarbons (PAHs; \citealt{Leger & Puget 1984}; \citealt{Allamandola et al. 1985}; see review in \citealt{Tielens 2008}) produce among the most prominent spectral features in the mid-infrared (IR) spectra of active and inactive galaxies. The main PAH features at 6.2, 7.7, 8.6, 11.3, and 12.7\,$\mum$ together can account for up to 20\% of the total IR emission in star-forming galaxies (SFGs; \citealt{Smith et al. 2007b}; \citealt{Xie et al. 2018}; \citealt{Li 2020}). PAHs radiate through IR fluorescence following vibrational excitation after absorbing a single ultraviolet (UV) photon (\citealt{Allamandola et al. 1989}; \citealt{Tielens 2005}). Therefore, the intensity of PAH emission indirectly traces the strength of the UV radiation field, and hence recent star formation (\citealt{Forster Schreiber et al. 2004}; \citealt{Peeters et al. 2004}).

A number of works have focused on calibrating the strength of individual PAH bands, or some combination thereof, against other traditionally well-established tracers of star formation, concluding that PAH can serve as an effective indicator of star formation rate (SFR) in different galaxy environments (e.g., \citealt{Calzetti et al. 2005, Calzetti et al. 2007};  \citealt{Wu et al. 2005}; \citealt{Treyer et al. 2010}; \citealt{Shipley et al. 2016}; \citealt{Maragkoudakis et al. 2018}; \citealt{Xie & Ho 2019}). Meanwhile, extensive observations have revealed that the relative intensity of individual PAH features varies greatly across different environments (e.g., \citealt{Genzel et al. 1998}; \citealt{Kaneda et al. 2005}; \citealt{Farrah et al. 2007}; \citealt{Gordon et al. 2008}; \citealt{ODowd et al. 2009}; \citealt{Hunt et al. 2010}; \citealt{Sales et al. 2010}; \citealt{Lebouteiller et al. 2011}). In particular, the harsh conditions around active galactic nuclei (AGNs) are hostile to the survival of PAHs because they can be eroded by extreme-UV or X-ray photons (\citealt{Aitken & Roche 1985}; \citealt{Voit 1992}).

Studies of the mid-IR spectra of nearby galaxies \citep{Smith et al. 2007b, Diamond-Stanic & Rieke 2010} reveal that low-luminosity AGNs (LLAGNs), such as Seyfert nuclei and low-ionization nuclear emission-line regions (LINERs; \citealt{Heckman 1980, Ho 2008}), tend to exhibit weaker 6--8\,$\mum$ PAH emission, which can be attributed to selective destruction of smaller PAH grains responsible for the shorter wavelength features by the hard radiation field of the AGN.  \cite{ODowd et al. 2009} found that the PAH characteristics also vary with the age of the stellar population, as galaxies with younger populations have stronger short-wavelength PAH bands than older galaxies. However, when matched in stellar age, AGNs and galaxies with composite sources of ionization still manifest weaker 7.7\,$\mum$ emission than quiescent galaxies, likely a consequence of the preferential depletion of small grains by shocks and/or X-rays from the AGN.  While in extragalactic H\,{\small II} regions \citep{Gordon et al. 2008} and low-metallicity dwarf galaxies \citep{Hunt et al. 2010} the PAH strength diminishes with increasing hardness of the radiation field, as traced by the [Ne\,{\small III}]\,15.5\,$\mum$/[Ne\,{\small II}]\,12.8\,$\mum$ ratio, the suppression of PAH in galaxies hosting AGNs is even more severe \citep{Smith et al. 2007b}, hinting that more than radiative effects are operating.  Support for the role of shocks also comes from the correlation between the reduction of 7.7\,$\mum$/11.3\,$\mum$ with increasing prominence of the shock-sensitive ${\rm H_{2}}\ S(3)$\,9.665\,$\mum$ rotational line \citep{Diamond-Stanic & Rieke 2010}.

The above considerations suggest that radiative effects by a hard radiation field alone cannot account for the PAH properties of AGNs. Some other factor, potentially shocks, may play a role.  Interstellar shocks with velocities $\gtrsim 100\, {\rm km\ s^{-1}}$ in principle can completely destroy large PAHs with more than $N_{\rm C} = 50$ carbon atoms through the collision between energetic ions/electrons and PAHs \citep{Micelotta et al. 2010a}. Moreover, PAHs immersed in the hot post-shock gas can also be destroyed by collisions with ions and electrons, with smaller PAHs suffering the fastest damage (\citealt{Micelotta et al. 2010b}). Mechanical processing, therefore, produces strong variations in the strength of the aromatic features, as PAHs of different sizes peak in emission efficiency at different wavelengths.  For instance, PAHs with $N_{\rm C} \simeq 50, 100$, and 500--1000 radiate their peak efficiency at 6.2, 7.7, and 11.3\,$\mum$, respectively \citep{Draine & Li 2007}. Despite the implementation of shocks into modern models of dust evolution (e.g., \citealt{Murga et al. 2016, Murga et al. 2019}; \citealt{Hirashita & Murga 2020}; \citealt{Hirashita et al. 2020}), observational evidence regarding the mechanical effect of shocks on PAH characteristics is still largely unestablished. 

The mixture of PAHs with different sizes or ionization states over multiple physical scales can lead to diverse PAH band ratios \citep{Draine & Li 2001, Draine & Li 2007, Draine et al. 2021}. Isolating the most AGN-dominated regions using spatially resolved analysis of mid-IR spectra can help to shed light on the underlying physical mechanisms responsible for the diversity of PAH emission in active galaxies. Very few such attempts have been made. In this paper, we perform a comprehensive investigation of the mid-IR PAH features using spatially resolved, mapping-mode observations obtained by the Spitzer Infrared Nearby Galaxies Survey (SINGS; \citealt{Kennicutt et al. 2003}) acquired with the Infrared Spectrograph (IRS; \citealt{Houck et al. 2004}) on the Spitzer Space Telescope (\citealt{Werner et al. 2004}).  Section~\ref{section:sec2} describes the observational material and data processing, including the methodology to measure PAH emission. The results of the spatially resolved analysis are presented in Section~\ref{section:sec3}. Section~\ref{section:sec4} discusses the physical mechanisms responsible for the differences in PAH band ratios in AGNs and SFGs, as well as implications for AGN feedback. Main conclusions are summarized in Section~\ref{section:sec5}.

\section{Observational Material}\label{section:sec2}

\subsection{Sample and Data} 

SINGS\footnote{\url{https://irsa.ipac.caltech.edu/data/SPITZER/SINGS/}} is a comprehensive IR imaging and spectroscopic survey of 75 nearby galaxies conducted using Spitzer (\citealt{SINGS 2020}).  To obtain spatially resolved mid-IR emission lines for the central regions of the galaxies, we utilize the three-dimensional spectral datacubes released by the survey team, which were constructed using the {\tt CUBISM} tool (\citealt{Smith et al. 2007a}).  The sample spans a representative range of galaxy properties, including nuclear activity, and hence is well suited to study the impact of AGN activity on PAH emission.  This paper focuses on the 66 galaxies that have successful IRS mapping-mode observations covering $\sim 5-38\ \mum$.  Table~\ref{tab:tablegal} provides a summary of a few key galaxy properties.    Nuclear spectral classifications are taken from \cite{Moustakas et al. 2010}, but preference is given to the more detailed classifications of \cite{Ho et al. 1997a, Ho et al. 1997b}, whenever available. Bolometric luminosities for the 30 AGNs in the sample are based on either total (narrow plus broad) H$\alpha$ emission \citep{Ho et al. 2003} or nuclear 2--10 keV X-ray emission \citep{Ho 2009}, using bolometric corrections appropriate for LLAGNs \citep{Ho 2008}.  We estimate black hole masses from the $M_{\rm BH}-\sigma_{*}$ relation, as updated by \cite{Greene et al. 2020} for all galaxy types, with the aid of central stellar velocity dispersions from \cite{Ho et al. 2009}.

\begin{figure}[!t]
\center{\includegraphics[width=1\linewidth]{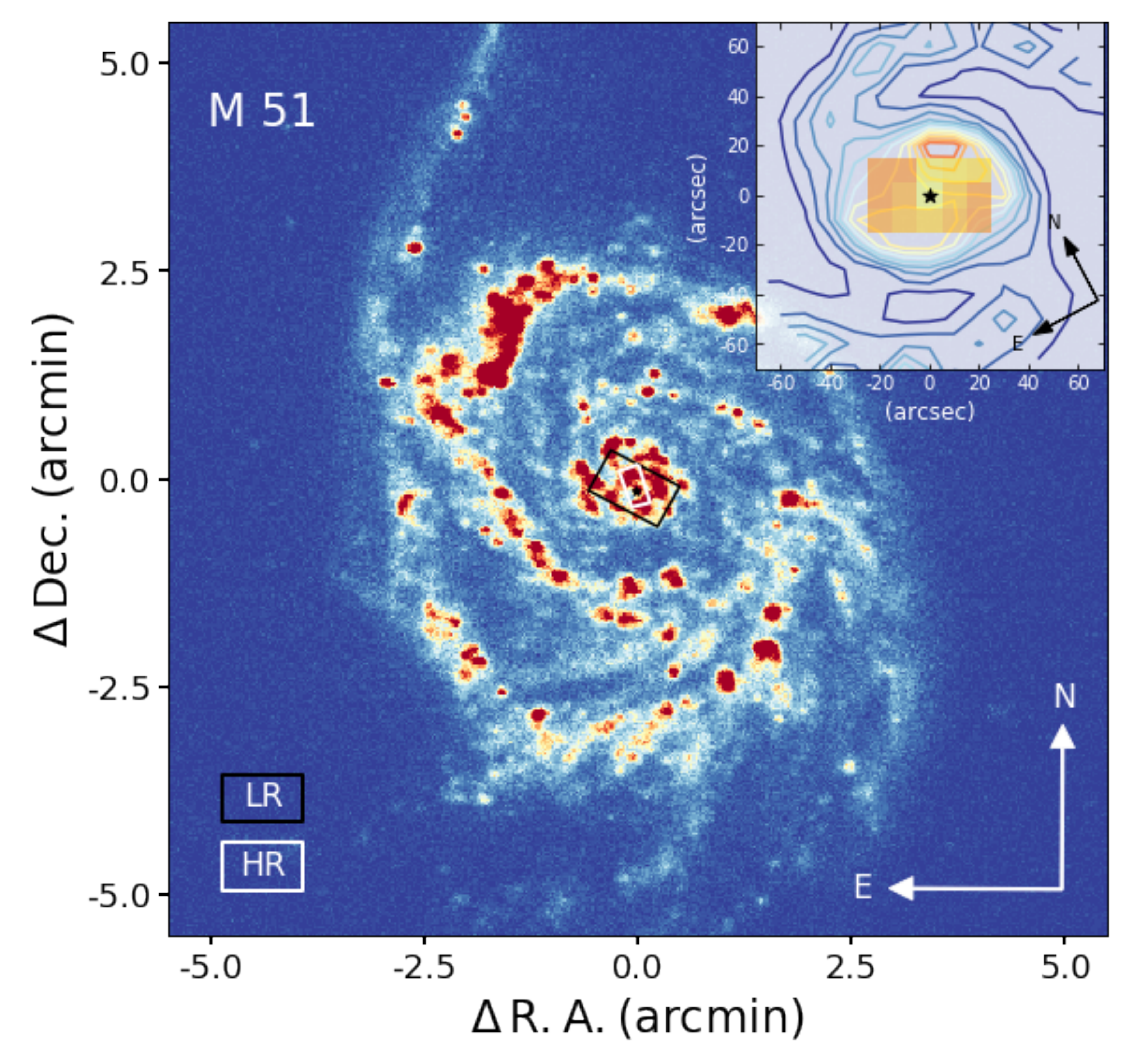}}
\caption{{Illustration of coverage of IRS mapping-mode observations of M\,51, overlaid on GALEX FUV image as the background. The black and white rectangles delineate the central region with complete SL+LL and SH+LH spectra, respectively, with the black star marking the center. The inset in the top-right corner is a zoom-in of the central $2\arcmin \times 2\arcmin$ region that shows the distribution of resolved $10\arcsec \times 10\arcsec$ spaxels within the black rectangle that has complete SL+LL spectra, color-coded (from orange to yellow) according to the flux of the IRS spectra integrated over $5 - 20\ \mum$.}\label{fig:coverage}}
\end{figure}

The IRS has both a low-resolution and a high-resolution module, each with two sets of slits to capture different wavelength ranges. The short-low (SL) mode has two orders with slit sizes of $3\farcs6 \times 57\arcsec$ (SL2) and $3\farcs7 \times 57\arcsec$ (SL1), covering, respectively, $5.2-7.7\ \mu$m and $7.4-14.5\ \mu$m, while the two orders of the long-low (LL) mode cover $14.0-21.3\ \mu$m with the $10\farcs5 \times 168\arcsec$ slit (LL2) and $19.5-38.0\ \mu$m with the $10\farcs7 \times 168\arcsec$ slit (LL1). The resolving power of the low-resolution module varies from $\lambda/\Delta\lambda\approx 64$ to 128. The short-high (SH) mode samples $9.9-19.6\ \mu$m with the $4\farcs7 \times 11\farcs3$ slit, while the long-high (LH) mode samples $18.7-37.2\ \mu$m  with the $11\farcs1 \times 22\farcs3$ slit, both with $\lambda/\Delta\lambda\approx 600$. The mapping-mode observations were conducted by scanning the slit to cover large sections of the galaxy. For the central region of each galaxy the mapping area covered $\sim 30\arcsec \times 50\arcsec$ with complete low-resolution (SL+LL) spectra and $\sim 15\arcsec \times 23\arcsec$ (nine galaxies have slightly larger coverage; \citealt{Dale et al. 2009}) with complete high-resolution (SH+LH) spectra. This paper focuses on the central regions covered by complete low-resolution mapping-mode observations. An example galaxy is shown in Figure~\ref{fig:coverage}.

To provide spatially resolved information on the star formation activity, we collect far-UV (FUV; 1350\,\AA--1750\,\AA) images from the Galaxy Evolution Explorer (GALEX) UV Atlas of Nearby Galaxies\footnote{\url{https://archive.stsci.edu/prepds/galex_atlas/}} (\citealt{Gil de Paz et al. 2007}).  To implement the methodology of \cite{Zhang et al. 2021} for spectral decomposition and to provide sufficient leverage to constrain the stellar continuum (Section~\ref{sec:section2.2}), we additionally make use of near-IR ($J, H, K_{s}$) images from the Two-Micron All Sky Survey (2MASS) Large Galaxy Atlas\footnote{\url{https://irsa.ipac.caltech.edu/data/LGA/}} \citep{Jarrett et al. 2003, Jarrett et al. 2020} and mid-IR (W1, W2, W3) images from WISE\footnote{\url{https://irsa.ipac.caltech.edu/applications/wise/}} \citep{Wright et al. 2010, IPAC 2020} and Spitzer (IRAC1, IRAC2, IRAC3, IRAC4, MIPS24). Table~\ref{tab:tableimg} lists the multiwavelength imaging datasets used in this paper. The IRAC4, W3, and MIPS24 bands also help to mitigate against the systematics in the convolution of the low-resolution IRS datacube (Section~\ref{sec:section2.2}). To correct for the differences in the photometric calibration between point sources and extended sources in the IRAC images, we adjust the IRAC flux density according to the recommendations in the IRAC Instrument Handbook.\footnote{\url{https://irsa.ipac.caltech.edu/data/SPITZER/docs/irac/iracinstrumenthandbook/29/}}

\startlongtable
\setlength{\tabcolsep}{2pt}
\begin{deluxetable*}{cccccll||cccccll}
\tabletypesize{\footnotesize}
\tablecolumns{14}
\tablecaption{Properties of the Galaxy Sample}
\tablehead{
\colhead{Galaxy} & \colhead{Hubble} & \colhead{Nuclear} & \colhead{$D_L$} & \colhead{\scriptsize$E(B-V)$} & \colhead{$\log L_{\rm bol}$} & \colhead{$\log M_{\rm BH}$} & \colhead{Galaxy} & \colhead{Hubble} & \colhead{Nuclear} & \colhead{$D_L$} & \colhead{\scriptsize$E(B-V)$}& \colhead{$\log L_{\rm bol}$} & \colhead{$\log M_{\rm BH}$}\\
\colhead{Name} & \colhead{Type} & \colhead{Classification} & \colhead{(Mpc)} & \colhead{(mag)} & \colhead{($\rm erg\ s^{-1}$)} & \colhead{($M_\odot$)} & \colhead{Name} & \colhead{Type} & \colhead{Classification} & \colhead{(Mpc)} &\colhead{(mag)} & \colhead{($\rm erg\ s^{-1}$)} & \colhead{($M_\odot$)}\\
\colhead{(1)} & \colhead{(2)} & \colhead{(3)} & \colhead{(4)} & \colhead{(5)} & \colhead{(6)} & \colhead{(7)} & \colhead{(1)} & \colhead{(2)} & \colhead{(3)} & \colhead{(4)} & \colhead{(5)} & \colhead{(6)} & \colhead{(7)} 
}
\startdata
DDO~53 & Im & SFG & 3.61 & 0.032 & \nodata & \nodata & NGC~3521 & SABbc & AGN(H/L2::) & 11.2 & 0.050 & 41.04\tablenotemark{{\scriptsize $\ g$}} & 7.50 \\
DDO~165 & Im & SFG & 4.57 & 0.020 & \nodata & \nodata & NGC~3621 & Sd & AGN & 6.55 & 0.069 & 40.49\tablenotemark{{\scriptsize $\ h \dag$}} & 6.48\tablenotemark{{\scriptsize $\ h \dag$}} \\
Holm~IX & Im & SFG & 3.5 & 0.067 & \nodata & \nodata & NGC~3627 & SABb & AGN(T2/S2) & 9.38 & 0.029 & 38.72 & 7.40 \\
IC~4710 & SBm & SFG & 8.5 & 0.077 & \nodata & \nodata & NGC~3773 & S0 & SFG & 12.4 & 0.023 & \nodata & \nodata \\
Mrk~33 & Im & SFG & 21.7 & 0.010 & \nodata & \nodata & NGC~3938 & Sc & SFG & 17.9 & 0.018 & \nodata & \nodata \\
NGC~24 & Sc & SFG & 8.2 & 0.017 & \nodata & \nodata & NGC~4125 & E6 & AGN(T2) & 21.4 & 0.016 & 40.03 & 8.54 \\
NGC~337 & SBd & SFG & 19.3 & 0.096 & \nodata & \nodata & NGC~4236 & SBdm & SFG & 4.45 & 0.013 & \nodata & \nodata \\
NGC~584 & E4 & AGN & 20.8 & 0.036 & 40.26\tablenotemark{{\scriptsize $\ a \dag$}} & 8.13\tablenotemark{{\scriptsize $\ a$}} & NGC~4254 & Sc & SFG & 14.4 & 0.033 & \nodata & \nodata \\
NGC~628 & Sc & SFG & 7.2 & 0.061 & \nodata & \nodata & NGC~4321 & SABbc & AGN(T2) & 14.3 & 0.023 & 39.68 & 6.64 \\
NGC~855 & E & SFG & 9.73 & 0.061 & \nodata & \nodata & NGC~4450 & Sab & AGN(L1.9) & 20.0 & 0.024 & 41.37 & 7.56 \\
NGC~925 & SABd & SFG & 9.12 & 0.066 & \nodata & \nodata & NGC~4536 & SABbc & SFG & 14.5 & 0.016 & \nodata & \nodata \\
NGC~1097 & SBb & AGN(S1) & 14.2 & 0.023 & 41.80\tablenotemark{{\scriptsize $\ b$}} & 8.10\tablenotemark{{\scriptsize $\ b$}} & NGC~4552 & E & AGN(T2:) & 4.9 & 0.035 & 39.62 & 8.74 \\
NGC~1266 & SB0 & AGN(L) & 30.6 & 0.085 & 43.55\tablenotemark{{\scriptsize $\ c \dag$}} & 6.23\tablenotemark{{\scriptsize $\ c$}} & NGC~4559 & SABcd & SFG & 6.98 & 0.015 & \nodata & \nodata \\
NGC~1291 & SB0/a & AGN & 10.4 & 0.011 & 40.56\tablenotemark{{\scriptsize $\ b$}} & 8.00\tablenotemark{{\scriptsize $\ b$}} & NGC~4569 & SABab & AGN(T2) & 9.86 & 0.040 & 40.15 & 7.57 \\
NGC~1316 & SAB0 & AGN(L) & 21.0 & 0.018 & 40.48\tablenotemark{{\scriptsize $\ d$}} & 8.44\tablenotemark{{\scriptsize $\ d$}} & NGC~4579 & SABb & AGN(S1.9/L1.9) & 16.4 & 0.035 & 42.33 & 7.94 \\
NGC~1377 & S0 & SFG & 24.6 & 0.024 & \nodata & \nodata & NGC~4594 & Sa & AGN(L2) & 9.08 & 0.044 & 41.20 & 8.65 \\
NGC~1404 & E1 & AGN & 20.2 & 0.010 & 41.73\tablenotemark{{\scriptsize $\ d$}} & 8.50\tablenotemark{{\scriptsize $\ d$}} & NGC~4625 & SABm & SFG & 9.3 & 0.016 & \nodata & \nodata \\
NGC~1482 & S0 & SFG & 22.6 & 0.034 & \nodata & \nodata & NGC~4631 & SBd & SFG & 7.62 & 0.015 & \nodata & \nodata \\
NGC~1512 & SBab & AGN & 11.6 & 0.009 & 40.37\tablenotemark{{\scriptsize $\ e \dag$}} & 7.13\tablenotemark{{\scriptsize $\ e$}} & NGC~4725 & SABab & AGN(S2:) & 11.9 & 0.010 & 40.27 & 7.63 \\
NGC~1566 & SABbc & AGN(S1.5) & 18.0 & 0.008 & 42.49\tablenotemark{{\scriptsize $\ f$}} & 6.83\tablenotemark{{\scriptsize $\ f$}} & NGC~4736 & Sab & AGN(L2) & 4.66 & 0.015 & 39.75 & 7.21 \\
NGC~1705 & Am & SFG & 5.8 & 0.007 & \nodata & \nodata & NGC~4826 & Sab & AGN(T2) & 5.27 & 0.036 & 38.75 & 6.92 \\
NGC~2403 & SABcd & SFG & 3.5 & 0.034 & \nodata & \nodata & NGC~5033 & Sc & AGN(S1.5) & 13.3 & 0.010 & 41.60 & 7.77 \\
NGC~2798 & SBa & SFG & 25.8 & 0.017 & \nodata & \nodata & NGC~5055 & Sbc & AGN(T2) & 7.94 & 0.015 & 39.65 & 7.29 \\
NGC~2841 & Sb & AGN(L2) & 14.1 & 0.013 & 39.60 & 8.50 & NGC~5194 & SABbc & AGN(S2) & 8.2 & 0.031 & 42.28 & 6.92 \\
NGC~2915 & I0 & SFG & 3.78 & 0.236 & \nodata & \nodata & NGC~5195 & SB0 & AGN(L2:) & 8.2 & 0.030 & 39.09 & 7.41 \\
NGC~2976 & Sc & SFG & 3.55 & 0.064 & \nodata & \nodata & NGC~5408 & IBm & SFG & 4.8 & 0.059 & \nodata & \nodata \\
NGC~3031 & Sab & AGN(S1.5) & 3.5 & 0.069 & 41.37 & 7.90 & NGC~5713 & SABbc & SFG & 21.4 & 0.034 & \nodata & \nodata \\
NGC~3049 & SBab & SFG & 19.2 & 0.033 & \nodata & \nodata & NGC~5866 & S0 & AGN(T2) & 15.3 & 0.011 & 39.80 & 7.98 \\
NGC~3184 & SABcd & SFG & 11.7 & 0.014 & \nodata & \nodata & NGC~6946 & SABcd & SFG & 6.8 & 0.294 & \nodata & \nodata \\
NGC~3190 & Sa & AGN(L2) & 19.3 & 0.022 & 40.61 & 8.18 & NGC~7331 & Sb & AGN(T2) & 14.5 & 0.078 & 39.87 & 7.59 \\
NGC~3198 & SBc & SFG & 14.1 & 0.011 & \nodata & \nodata & NGC~7552 & Sc & SFG & 22.3 & 0.012 & \nodata & \nodata \\
NGC~3265 & E & SFG & 19.6 & 0.021 & \nodata & \nodata & NGC~7793 & Sd & SFG & 3.91 & 0.017 & \nodata & \nodata \\
NGC~3351 & SBb & SFG & 9.33 & 0.024 & \nodata & \nodata & Tol~89 & SBdm & SFG & 7.66 & 0.057 & \nodata & \nodata \\
\enddata
\tablecomments{Col. (1): Galaxy name.  Col. (2): Hubble type taken from NASA/IPAC Extragalactic Database (NED). Col. (3): Nuclear classification according to the [O\,III]/H$\beta$ versus [N\,II]/H$\alpha$ emission-line diagnostic diagram based on the spectra extracted from a central aperture of $2\farcs5 \times 2\farcs5$, which distinguish it as a star-forming galaxy (SFG) or an AGN (see \citealt{Moustakas et al. 2010} for details). We provide more specific classifications for the AGN if available from \cite{Ho et al. 1997a, Ho et al. 1997b} or NED, where S = Seyfert, L = LINER, T = transition object, 1 = type~1 (broad-line), 2 = type~2 (narrow-line), a number between 1 and 2 pertains to an intermediate-type, and ``:'' indicates an uncertain classification \citep{Ho et al. 1997a}. Col. (4): Luminosity distance taken from the updated SINGS photometry catalog \citep{Dale et al. 2017}. Col. (5): Galactic reddening from Schlafly \& Finkbeiner (2011).  Col. (6): Bolometric luminosity calculated based on the nuclear 2--10 keV luminosity, with luminosity distance adjusted to that adopted here, and bolometric correction $C_{\rm X} = 15.8$ (\citealt{Ho 2009}), unless otherwise specified. Col. (7): Black hole mass derived from the $M_{\rm BH}-\sigma_{*}$ relation (for all galaxy types) from \cite{Greene et al. 2020} and the central stellar velocity dispersion measured by \cite{Ho et al. 2009}, unless otherwise specified.}

\tablenotetext{{\scriptsize a-g}}{\ \ \ \ \ \ \ \ References for $L_{\rm bol}$ and $M_{\rm BH}$: (a) \citet[converted to 2--10 keV luminosity based on $\Gamma$ and $N_{\rm H}$ therein]{Lehmer et al. 2019} and \cite{Thater et al. 2019}, respectively; (b) \citet[and references therein]{Cisternas et al. 2013}; (c) \citet[$L_{\rm bol}$ based on IR SED fitting]{Alatalo et al. 2015}; (d) \citet[and references therein]{Pellegrini 2010}; (e) \cite{Ducci et al. 2014} and \cite{Dullo et al. 2020}, respectively; (f) \cite{Combes et al. 2019}; (g) $L_{\rm bol}$ based on nuclear H$\alpha$ emission from \cite{Ho et al. 1997a}, as updated in \cite{Ho et al. 2003}, with bolometric correction factor $C_{\rm H\alpha} = 300$ (\citealt{Ho 2009}); (h) \cite{Gliozzi et al. 2009} and \cite{Barth et al. 2009}, respectively.}
\tablenotetext{{\scriptsize \dag}}{Upper limit.}
\label{tab:tablegal}
\end{deluxetable*}

\begin{deluxetable}{cccccc}[!ht]
\tabletypesize{\footnotesize}
\tablecolumns{6}
\tablecaption{Summary of Multiwavelength Imaging Data}
\tablehead{
\colhead{Telescope} & \colhead{Filter} & \colhead{$\lambda_{\rm eff}$} & \colhead{FWHM} & \colhead{$\sigma_{\rm cal}$} & \colhead{References}\\
\colhead{} & \colhead{} & \colhead{$(\mu m)$} & \colhead{(\arcsec)} & \colhead{(\%)} & \colhead{}\\
\colhead{(1)} & \colhead{(2)} & \colhead{(3)} & \colhead{(4)} & \colhead{(5)} & \colhead{(6)}
}
\startdata
GALEX & FUV & 0.153 & 4.2 & 5 & 1,2\\
2MASS & $J$  & 1.25 & 2.5 & 5 & 3, 4\\
2MASS & $H$ & 1.65 & 2.5 & 5 & 3, 4\\
2MASS & $K_s$ & 2.16 & 2.5 & 5 & 3, 4\\
WISE &  W1 & 3.35 & 6.1 & 2.4 & 5, 6\\
WISE &  W2 & 4.60 & 6.4 & 2.8 & 5, 6\\
WISE &  W3 & 11.56 & 6.5 & 4.5 & 5, 6\\
Spitzer & IRAC1 & 3.55 & 1.7 & 10 & 7, 8\\
Spitzer & IRAC2 & 4.49 & 1.7 & 10 & 7, 8\\
Spitzer & IRAC3 & 5.73 & 1.9 & 10 & 7, 8\\
Spitzer & IRAC4 & 7.87 & 2.0 & 10 & 7, 8\\
Spitzer & MIPS24 & 23.68 & 6.0 & 4 & 7, 9\\
\enddata
\tablecomments{Col. (1): Telescope. Col. (2): Filter. Col. (3): Effective wavelength of the filter. Col. (4): FWHM of the PSF. Col. (5): Calibration uncertainty.  Col. (6):  References.}
\tablerefs{(1) \citealt{Morrissey et al. 2007}; (2) \citealt{Gil de Paz et al. 2007}; (3) \citealt{Jarrett et al. 2003}; (4) \citealt{Skrutskie et al. 2006}; (5) \citealt{Wright et al. 2010}; (6) \citealt{Jarrett et al. 2013}; (7) \citealt{Dale et al. 2007}; (8) \citealt{Fazio et al. 2004}; (9) \citealt{Engelbracht et al. 2007}.}
\label{tab:tableimg}
\end{deluxetable}

\subsection{Data Processing and Spectral Measurements}\label{sec:section2.2}

\subsubsection{Convolution, Optimal Binning, Spectral Extraction}\label{sec:section2.2.1}

To extract spatially resolved PAH emission, we follow the strategy of \cite{Zhang et al. 2021}, which is summarize here.  Although the publicly released imaging and spectroscopic data of SINGS are already background-subtracted\footnote{\url{https://irsa.ipac.caltech.edu/data/SPITZER/SINGS/doc/sings_fifth_delivery_v2.pdf}}, we discovered that some ancillary images have evident background residuals that need further treatment. After masking out real sources with elliptical apertures through pipeline determination and visual check, we remove background residuals by fitting and subtracting a two-dimensional polynomial function. Then, both the images and spectral datacube are convolved to same angular resolution with ${\rm FWHM} = 10\arcsec$, which is comparable to but slightly broader than the coarsest PSF (${\rm FWHM} = 9\arcsec$) among all the images and slices of the spectral datacube (${\rm FWHM} \approx 1\farcs5-3\farcs5$ for the SL mode to $4\farcs5-9\farcs0$ for the LL mode). After the convolution, we reproject all the data into the same coordinate frame using a final pixel size of 10\arcsec, which corresponds to a physical scale of $\sim 0.17 - 1.25$~kpc for the galaxies studied here and ensures that all the spaxels within each galaxy are spatially independent. In the following analysis, the spatially resolved spaxels from each galaxy are regarded as independent samplings of similar galactic environments with different PAH characteristics, and, unless otherwise stated, we do not discriminate among the resolved spaxels from different galaxies.

\begin{figure}[!ht]
\center{\includegraphics[width=1\linewidth]{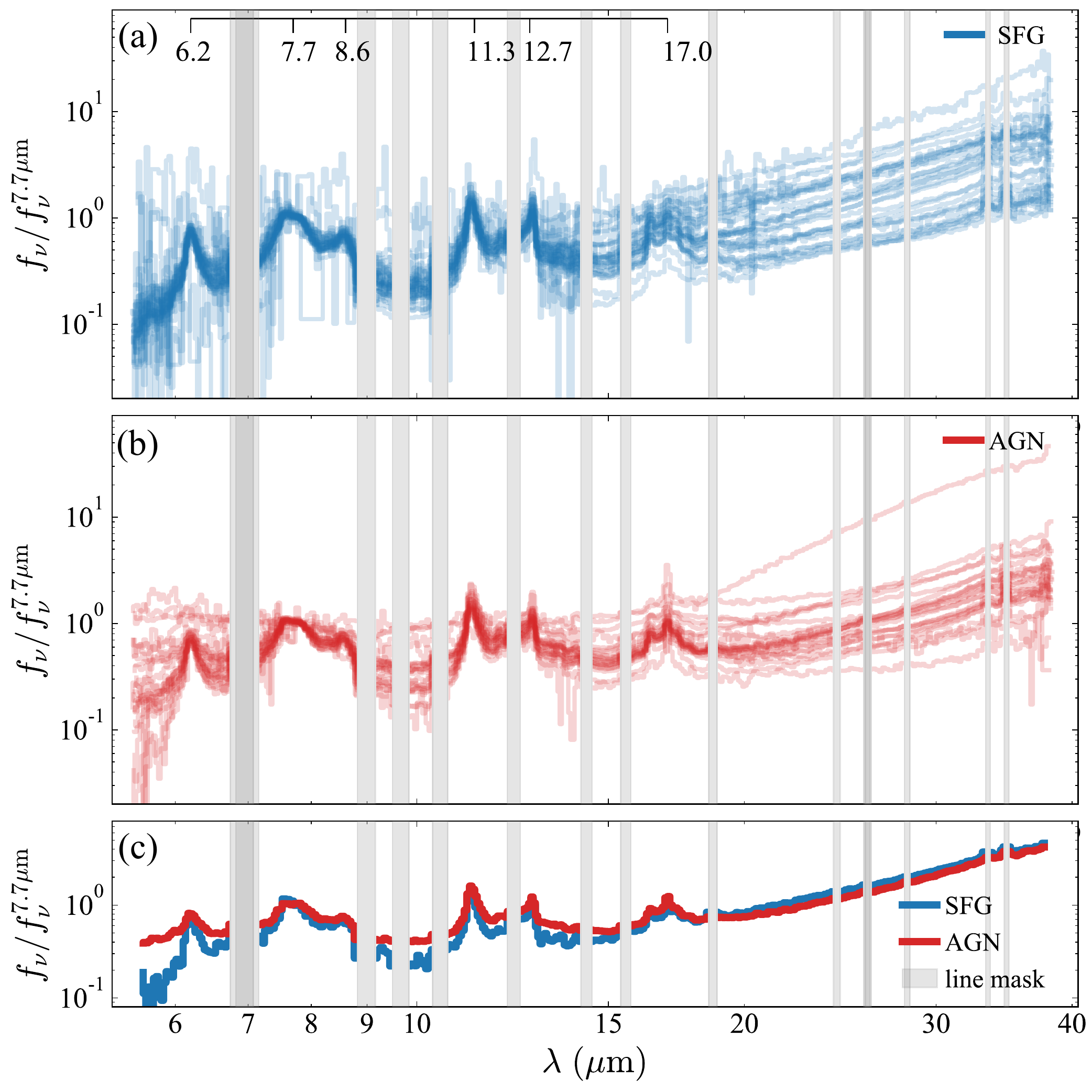}}
\caption{{Integrated low-resolution IRS spectra, with ionic emission lines masked, for the central region of galaxies classified as (a) SFGs and (b) AGNs. The main PAH features are marked in panel (a). Each integrated spectrum is obtained by summing the spectra of all the spaxels within the corresponding central region and normalized by the flux density at 7.7\,$\mum$. Panel (c) shows the composite spectra for the regions from SFGs (blue) and AGNs (red).}\label{fig:irs_spc}}
\end{figure}

We optimally bin the spectral datacube into final spaxels that simultaneously maximize spatial resolution while ensuring sufficient signal-to-noise ratio (S/N) to yield accurate PAH measurements \citep{Zhang et al. 2021}. The binning process considers the similarity between the spectra of different pixels to maximize the association of spatial structures in physical properties. The ancillary images are binned to exactly the same corresponding spaxel scale to facilitate the construction of spatially resolved spectral energy distributions (SEDs).  The resolved photometry for the 12 photometric bands is provided in Table~\ref{tab:TablePhotI}.  Finally, to mitigate against systematics in the convolution of the spectral datacube, the IRS spectrum of each binned spaxel is scaled by a numerical factor determined by minimizing the weighted difference between the observed flux densities and the flux densities synthesized from the IRS spectrum using the response curves of the mid-IR photometric bands. The above procedure generates 1 to 15 low-resolution spectra ($\sim 5-38\ \mum$) and associated spatially resolved SEDs for the central region of each galaxy (see an example in the inset of Figure~\ref{fig:coverage}). Figure~\ref{fig:irs_spc} illustrates the collection of integrated low-resolution spectra for the central region of all SFGs and AGNs studied in this paper, as well as their respective average composite spectra.

\startlongtable
\setlength{\tabcolsep}{2pt}
\begin{deluxetable*}{lcccccccccccc}
\tabletypesize{\scriptsize}
\tablecolumns{13}
\tablecaption{Spatially Resolved Multi-band Photometry}
\tablehead{
\colhead{Region} & \colhead{log $f_{\rm FUV}$} & \colhead{log $f_{J}$} & \colhead{log $f_{H}$} & \colhead{log $f_{Ks}$} & \colhead{log $f_{\rm W1}$} & \colhead{log $f_{\rm W2}$} & \colhead{log $f_{\rm W3}$} & \colhead{log $f_{\rm IRAC1}$} & \colhead{log $f_{\rm IRAC2}$} & \colhead{log $f_{\rm IRAC3}$} & \colhead{log $f_{\rm IRAC4}$} & \colhead{log $f_{\rm MIPS24}$}\\
\colhead{} & \colhead{(mJy)} &  \colhead{(mJy)} & \colhead{(mJy)} & \colhead{(mJy)} & \colhead{(mJy)} & \colhead{(mJy)} & \colhead{(mJy)} & \colhead{(mJy)} & \colhead{(mJy)} & \colhead{(mJy)} & \colhead{(mJy)} & \colhead{(mJy)}\\
\colhead{(1)} & \colhead{(2)} & \colhead{(3)} & \colhead{(4)} & \colhead{(5)} & \colhead{(6)} & \colhead{(7)} & \colhead{(8)} & \colhead{(9)} & \colhead{(10)} & \colhead{(11)} & \colhead{(12)} & \colhead{(13)}
}
\startdata
IC4710\_1 & $-$0.12 $\pm$ 0.01 & 0.35 $\pm$ 1.37 & \nodata & \nodata & 0.58 $\pm$ 0.02 & 0.33 $\pm$ 0.03 & 0.78 $\pm$ 0.01 & 0.53 $\pm$ 0.12 & 0.36 $\pm$ 0.20 & $-$0.01 $\pm$ 0.33 & 0.71 $\pm$ 0.05 & 1.34 $\pm$ 0.02 \\
Mrk33\_1 & 0.25 $\pm$ 0.02 & 1.06 $\pm$ 0.03 & 1.12 $\pm$ 0.03 & 1.05 $\pm$ 0.03 & 0.81 $\pm$ 0.01 & 0.66 $\pm$ 0.01 & 1.66 $\pm$ 0.02 & 0.90 $\pm$ 0.04 & 0.78 $\pm$ 0.04 & 1.32 $\pm$ 0.04 & 1.73 $\pm$ 0.04 & 2.65 $\pm$ 0.02 \\
Mrk33\_2 & $-$0.15 $\pm$ 0.02 & 0.66 $\pm$ 0.04 & 0.73 $\pm$ 0.05 & 0.65 $\pm$ 0.06 & 0.52 $\pm$ 0.01 & 0.38 $\pm$ 0.01 & 1.38 $\pm$ 0.02 & 0.54 $\pm$ 0.05 & 0.43 $\pm$ 0.04 & 0.95 $\pm$ 0.04 & 1.36 $\pm$ 0.04 & 2.26 $\pm$ 0.02 \\
Mrk33\_3 & $-$0.49 $\pm$ 0.02 & 0.58 $\pm$ 0.05 & 0.66 $\pm$ 0.06 & 0.56 $\pm$ 0.08 & 0.42 $\pm$ 0.01 & 0.27 $\pm$ 0.01 & 1.29 $\pm$ 0.02 & 0.38 $\pm$ 0.05 & 0.26 $\pm$ 0.05 & 0.79 $\pm$ 0.04 & 1.18 $\pm$ 0.04 & 1.97 $\pm$ 0.02 \\
Mrk33\_4 & $-$0.39 $\pm$ 0.02 & 0.61 $\pm$ 0.05 & 0.66 $\pm$ 0.06 & 0.56 $\pm$ 0.08 & 0.36 $\pm$ 0.01 & 0.19 $\pm$ 0.01 & 1.16 $\pm$ 0.02 & 0.36 $\pm$ 0.05 & 0.21 $\pm$ 0.05 & 0.67 $\pm$ 0.05 & 1.06 $\pm$ 0.04 & 1.80 $\pm$ 0.02 \\
Mrk33\_5 & $-$0.32 $\pm$ 0.01 & 0.95 $\pm$ 0.05 & 1.00 $\pm$ 0.07 & 0.90 $\pm$ 0.09 & 0.76 $\pm$ 0.01 & 0.60 $\pm$ 0.01 & 1.59 $\pm$ 0.01 & 0.67 $\pm$ 0.04 & 0.53 $\pm$ 0.03 & 0.93 $\pm$ 0.03 & 1.32 $\pm$ 0.02 & 1.91 $\pm$ 0.01 \\
Mrk33\_6 & $-$2.22 $\pm$ 0.35 & 0.03 $\pm$ 0.26 & 0.17 $\pm$ 0.30 & 0.03 $\pm$ 0.42 & $-$0.17 $\pm$ 0.01 & $-$0.37 $\pm$ 0.02 & 0.59 $\pm$ 0.01 & $-$0.31 $\pm$ 0.18 & $-$0.46 $\pm$ 0.10 & $-$0.76 $\pm$ 0.52 & 0.11 $\pm$ 0.07 & 0.23 $\pm$ 0.07 \\
NGC24\_1 & 0.08 $\pm$ 0.01 & 1.71 $\pm$ 0.01 & 1.77 $\pm$ 0.02 & 1.66 $\pm$ 0.03 & 1.41 $\pm$ 0.00 & 1.17 $\pm$ 0.00 & 1.41 $\pm$ 0.01 & 1.38 $\pm$ 0.02 & 1.20 $\pm$ 0.02 & 1.29 $\pm$ 0.11 & 1.52 $\pm$ 0.03 & 1.50 $\pm$ 0.01 \\
NGC337\_1 & $-$0.62 $\pm$ 0.02 & 0.62 $\pm$ 0.06 & 0.69 $\pm$ 0.06 & 0.61 $\pm$ 0.10 & 0.51 $\pm$ 0.01 & 0.39 $\pm$ 0.01 & 1.26 $\pm$ 0.02 & 0.57 $\pm$ 0.04 & 0.47 $\pm$ 0.04 & 1.05 $\pm$ 0.04 & 1.45 $\pm$ 0.04 & 2.05 $\pm$ 0.02 \\
NGC337\_2 & $-$0.78 $\pm$ 0.03 & 0.93 $\pm$ 0.04 & 1.01 $\pm$ 0.04 & 0.92 $\pm$ 0.05 & 0.70 $\pm$ 0.01 & 0.51 $\pm$ 0.01 & 1.20 $\pm$ 0.02 & 0.72 $\pm$ 0.04 & 0.57 $\pm$ 0.04 & 1.00 $\pm$ 0.04 & 1.37 $\pm$ 0.04 & 1.63 $\pm$ 0.02 \\
\enddata
\tablecomments{\footnotesize Spatially resolved multi-band flux densities and corresponding uncertainties.  (This table is available in its entirety in machine-readable form.)}
\label{tab:TablePhotI}
\end{deluxetable*}

The physical diagnostics in the following analysis (e.g., hardness of the radiation field, ionization parameter, and shock strength) are based on the narrow ionic and molecular emission lines.  Although these features can be accessed from the low-resolution spectra, their measurement is highly sensitive to the treatment of the underlying PAH and continuum emission, and in practice we can measure them reliably only from the high-resolution spectra. The high-resolution spectra cover a much smaller area than that mapped in low-resolution mode (Figure~\ref{fig:coverage}), as a consequence of which we do not have matching high-resolution diagnostics for every spaxel with low-resolution spectra. Consequently, in our subsequent analysis (e.g., Sections~3 and 4), we resort to applying to all the spatially resolved low-resolution spaxels from the same galaxy a single value of the physical diagnostics inferred from the effectively spatially unresolved high-resolution spectra.  To provide a very rough validation of our approach, we extract from each high-resolution datacube spectra of two spatially separate sub-regions: an ``inner'' sub-region that covers the centralmost $10\arcsec \times 10\arcsec$ area, which corresponds to the central spaxel of the low-resolution datacube after projection, and another ``outer'' sub-region covering all the remaining area with high-resolution observations.  No convolution is applied to the high-resolution datacube, and the SH spectra are scaled to match the continuum level of the LH spectra.  These spectra will be discussed in Section~\ref{sec:section4.3}. 

\begin{figure}[!ht]
\center{\includegraphics[width=1\linewidth]{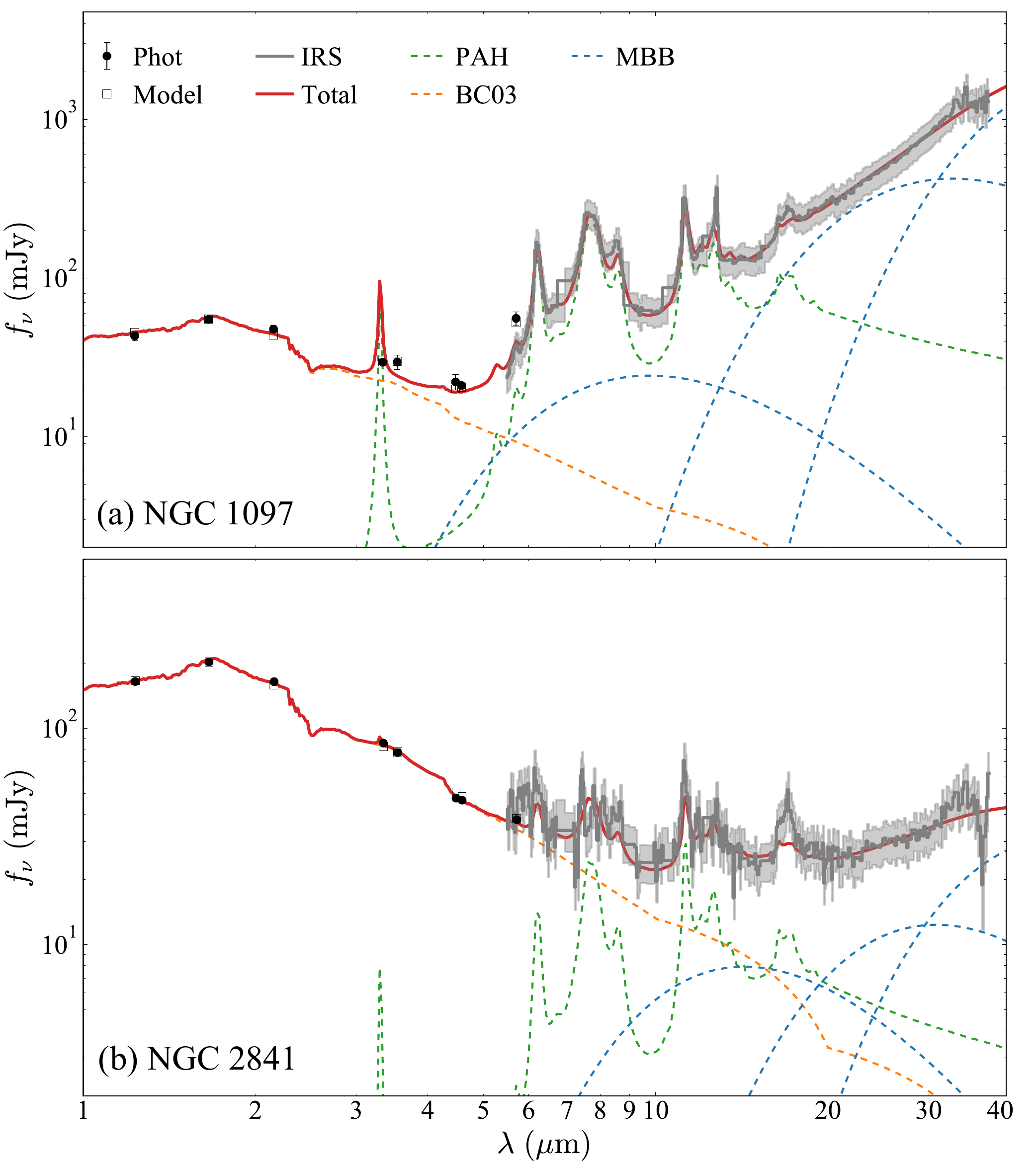}}
\caption{{Illustration of PAH decomposition for two sample IRS spectra, from (a) NGC~1097 and (b) NGC~2841. The gray solid line and black points are the observed IRS spectra and near-IR photometry. The red solid lines gives the best-fit model, which is composed of a PAH template (green dashed line), a stellar continuum component (orange dashed line) from \citet[BC03]{Bruzual & Charlot 2003}, and three dust continuum components (blue dashed lines), each represented by a modified blackbody (MBB).}\label{fig:SED_Fitting}}
\end{figure}

\begin{figure}[!ht]
\center{\includegraphics[width=1\linewidth]{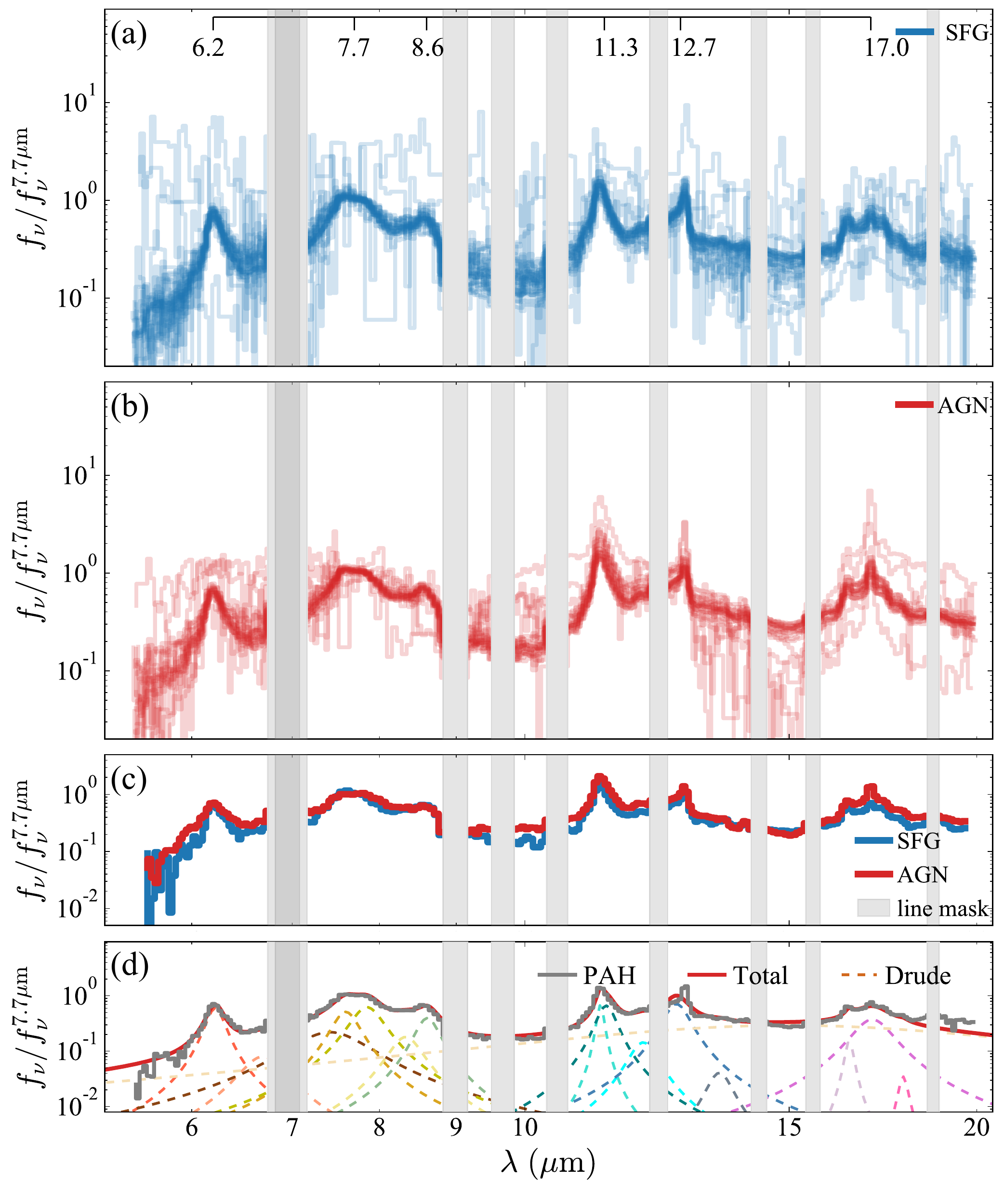}}
\caption{{Integrated residual PAH spectra for the central regions of (a) SFGs and (b) AGNs, obtained by summing the residual PAH spectra of all the spaxels within corresponding central region and normalized by the flux density at 7.7\,$\mum$. The residual PAH spectrum is produced by subtracting the best-fit continuum components from the observed spectrum. The main PAH features are marked in panel (a).  Panel (c) shows the composite residual PAH spectra for regions from SFGs (blue) and AGNs (red).  Panel (d) illustrates an example of decomposition of individual PAH features from the residual spectrum (grey line) using Drude profiles (dashed curves), which combine to produce the total model (red curve).}\label{fig:pah_spc}}
\end{figure}

\subsubsection{Measuring the PAH Features}\label{sec:section2.2.2}

Our spectral decomposition method for PAH measurement was designed and extensively tested for low-resolution IRS spectra of diverse galactic environments, including high-latitude clouds in the Milky Way, galaxies of ordinary to extreme degrees of star formation, as well as a wide range of levels of nuclear activity, from LLAGNs to systems powerful enough to qualify as quasars \citep{Xie et al. 2018}. This template fitting method, which can properly separate the PAH emission from the underlying continuum with only a few free parameters, was later applied to an even more extensive sample of SFGs covering broad galaxy properties \citep{Xie & Ho 2019}. \cite{Zhang et al. 2021} extended the method to spatially resolved, mapping-mode observations within individual galaxies, crucially by using supplementary photometry to treat the important regime when the IRS spectral coverage is incomplete (only SL or LL spectra are available).

After masking narrow emission lines that are not blended with the main PAH features, we fit the $\sim 5-38\, \mum$ spectrum with a multi-component model comprising a theoretical PAH template and three modified blackbodies of different temperatures to represent the dust continuum, all subject to attenuation by foreground extinction. In anticipation of other applications of the current dataset, some of which are explored in other papers in this series (e.g., \citealt{Zhang & Ho 2022}), we augment the fitting method by including an additional spectral component to account for the underlying stellar continuum at near-IR wavelengths.  We represent the starlight continuum using a stellar population model from \cite{Bruzual & Charlot 2003}, which assumes solar metallicity, a \cite{Chabrier 2003} stellar initial mass function, a constant star formation history, and a fixed stellar age of 6~Gyr.  This relatively simple model suffices for our purposes, as our primary focus is in the near-IR, which is mainly sensitive to evolved stars. We have verified that using more complicated star formation histories and a wider mixture of stellar ages makes little difference to our final results.

\begin{figure}[!ht]
\center{\includegraphics[width=1\linewidth]{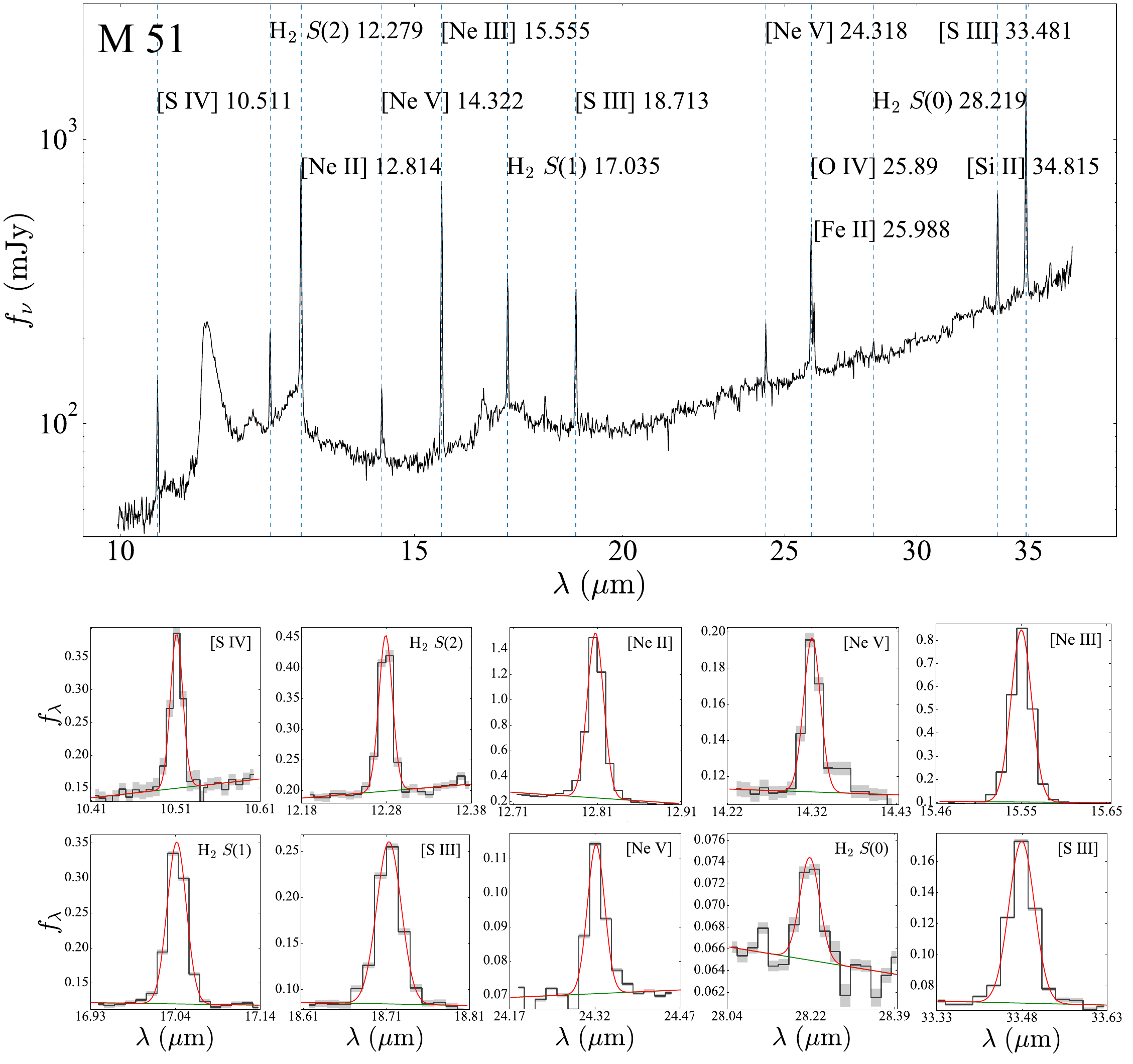}}
\caption{{(Top) Illustration of high-resolution IRS spectrum of M~51, marked with prominent ionic and molecular hydrogen emission lines. (Bottom) Zoom-in of individual narrow emission lines used in this paper (in units of $\rm 10^{-12}\ erg\ s^{-1}\ cm^{-2}\ \mum^{-1}$); the data are plotted as black histograms, with uncertainties indicated in gray. The red curve is the best-fitting Gaussian profile on top of a linear local continuum (green curve).}\label{fig:hr_spc}}
\end{figure}

\startlongtable
\setlength{\tabcolsep}{5pt}
\begin{deluxetable}{lcccc}
\tabletypesize{\scriptsize}
\tablecolumns{5}
\tablecaption{Spatially Resolved Measurements of PAH Emission}
\tablehead{
\colhead{Region} & \colhead{log $L_{\rm PAH}^{5-20}$} & \colhead{log $L_{\rm PAH}^{6.2}$} & \colhead{log $L_{\rm PAH}^{7.7}$} & \colhead{log $L_{\rm PAH}^{11.3}$} \\
\colhead{} & \colhead{($\rm erg\ s^{-1}$)} & \colhead{($\rm erg\ s^{-1}$)} & \colhead{($\rm erg\ s^{-1}$)} & \colhead{($\rm erg\ s^{-1}$)} \\
\colhead{(1)} & \colhead{(2)} & \colhead{(3)} & \colhead{(4)} & \colhead{(5)}}
\startdata
IC4710\_1 & 39.45 $\pm$ 0.09 & \nodata & \nodata & \nodata \\
Mrk33\_1 & 41.83 $\pm$ 0.02 & 40.90 $\pm$ 0.04 & 41.39 $\pm$ 0.04 & 40.84 $\pm$ 0.05 \\
Mrk33\_2 & 41.42 $\pm$ 0.02 & 40.40 $\pm$ 0.05 & 40.97 $\pm$ 0.05 & 40.37 $\pm$ 0.06 \\
Mrk33\_3 & 41.31 $\pm$ 0.02 & 40.36 $\pm$ 0.04 & 40.89 $\pm$ 0.03 & 40.29 $\pm$ 0.07 \\
Mrk33\_4 & 41.28 $\pm$ 0.01 & 40.33 $\pm$ 0.05 & 40.82 $\pm$ 0.05 & 40.25 $\pm$ 0.07 \\
Mrk33\_5 & 41.47 $\pm$ 0.02 & 40.50 $\pm$ 0.04 & 41.10 $\pm$ 0.03 & 40.45 $\pm$ 0.08 \\
Mrk33\_6 & 39.98 $\pm$ 0.05 & \nodata & \nodata & \nodata \\
NGC24\_1 & 40.61 $\pm$ 0.02 & 39.70 $\pm$ 0.04 & 40.27 $\pm$ 0.03 & 39.81 $\pm$ 0.04 \\
NGC337\_1 & 41.53 $\pm$ 0.02 & 40.57 $\pm$ 0.05 & 41.05 $\pm$ 0.06 & 40.44 $\pm$ 0.04 \\
NGC337\_2 & 41.38 $\pm$ 0.01 & 40.44 $\pm$ 0.04 & 40.95 $\pm$ 0.04 & 40.34 $\pm$ 0.04 \\
\enddata
\tablecomments{\footnotesize Col. (1): Resolved spaxels. Col. (2): Integrated PAH luminosity from $5 - 20\ \mum$ of best-fit PAH template. Cols. (3)--(5): PAH luminosity of PAH 6.2\,$\mum$, 7.7\,$\mum$ (7.414\,$\mum$, 7.598\,$\mum$, and 7.85\,$\mum$), and 11.3\,$\mum$ (11.23\,$\mum$ and 11.33\,$\mum$) features. (This table is available in its entirety in machine-readable form.) }
\label{tab:TablePAH}
\end{deluxetable}

Figure~\ref{fig:SED_Fitting} illustrates our multi-component spectral decomposition method for two example spectra, one representing a fairly typical case with prominent PAH emission and a strong dust continuum, and another for which the stellar component is strong but the PAH and dust emission are weak. The SED fitting is carried out using the Bayesian Markov chain Monte Carlo (MCMC) procedure {\tt emcee} in the {\tt Python} package.  As described in detail in \cite{Shangguan et al. 2018}, we include an extra 20\% uncertainty to the IRS spectrum in order to balance properly the relative weights of the sparse photometric points with the densely sampled spectroscopic data. We take the median and standard deviation of the posterior distribution of each best-fit parameter as the final estimate and its corresponding uncertainty. Based on the best-fit model for each spaxel, we define the integrated PAH luminosity, $L_{\rm PAH}$, as the integral of the best-fit PAH template over the $5 - 20\ \mum$ region.

The strengths of the individual PAH features are measured from the residual PAH spectrum by subtracting the best-fit stellar and dust continuum components from the observed spectrum with the extinction corrected. Figure~\ref{fig:pah_spc} displays the integrated residual PAH spectra for the central region of each galaxy, as well as the averaged, composite residual PAH spectra for galaxies classified as SFGs and AGNs (Figure~\ref{fig:pah_spc}c). We decompose the individual PAH features using a Bayesian MCMC procedure that fits a series of Drude profiles \citep{Draine & Li 2007} to the emission peaks (Figure~\ref{fig:pah_spc}d). \cite{Xie et al. 2018} showed that the measurements of individual PAH features based on our method agree well with those from other methods (e.g., {\tt PAHFIT}; \citealt{Smith et al. 2007b}). This paper focuses on the most prominent 6.2\,$\mum$ feature, the 7.7\,$\mum$ complex, and the 11.3\,$\mum$ complex, which have been measured for 453 spaxels (Table~\ref{tab:TablePAH}).

\subsubsection{Measuring the Narrow Emission Lines}\label{sec:section2.2.3}

The high-resolution spectra give access to a set of narrow ionic fine-structure lines and molecular hydrogen vibrational lines, which provide valuable diagnostics of the physical conditions of the interstellar medium (Figure~\ref{fig:hr_spc}, top panel).  The narrow emission lines have a simple profile at the current instrumental resolution, and we measure them by fitting a single Gaussian function plus a local, linear continuum (Figure~\ref{fig:hr_spc}, bottom panels). In addition to the spectrum integrated over the entire mapped area, we also have spectra extracted for the ``inner'' and ``outer'' sub-regions, as defined in Section~4.2. All the line diagnostics are based on extinction-corrected values, using the best-fit extinction derived from the low-resolution spectral fitting. Table~\ref{tab:FluxRatioI} lists measurements for all three extractions, emphasizing the following line-intensity ratios: [Ne\,{\small III}]\,15.5\,$\mum$/[Ne\,{\small II}]\,12.8\,$\mum$, an indicator of the hardness of the radiation field \citep{Thornley et al. 2000}, [S\,{\small IV}]\,10.5\,$\mum$/[Ne\,{\small III}]\,15.5\,$\mum$, a surrogate for the intensity of the radiation field (\citealt{Pereira-Santaella et al. 2017}), H$_{2}\,S(2)$\,12.3\,$\mum$/H$_{2}\,S(0)$\,28.2\,$\mum$, an indicator of gas temperature (\citealt{Turner et al. 1977}; \citealt{Vega et al. 2010}), and [S\,{\small III}]\,18.7\,$\mum$/[S\,{\small III}]\,33.5\,$\mum$, which is sensitive to electron density (\citealt{Dale et al. 2006}). Moreover, the intensity of H$_{2}\,S(0,1,2)$ relative to that of PAH gives an estimate of shock strength (\citealt{Roussel et al. 2007}).

\startlongtable
\begin{deluxetable*}{lccccccccccccccc}
\setlength{\tabcolsep}{2pt}
\tabletypesize{\scriptsize}
\tablecolumns{16}
\tablecaption{Mid-IR Diagnostic Line Ratios}
\tablehead{
\colhead{Galaxy} & & \colhead{log $\rm\frac{[Ne\ III]}{[Ne\ II]}$} & & & \colhead{log $\rm\frac{[S\ IV]}{[Ne\ III]}$} & & & \colhead{log $\rm\frac{[S\ III]\,18.7}{[S\ III]\,33.5}$} & & & \colhead{log $\frac{{\rm H_{2}}\ S(2)}{{\rm H_{2}}\ S(0)}$} & & & \colhead{log $\frac{{\rm H_{2}}\ S(0,1,2)}{{\rm PAH\ 7.7}}$} &\\
\cline{3-3}\cline{6-6}\cline{9-9}\cline{12-12}\cline{15-15}
\colhead{} & \colhead{\scriptsize \ \ \ \ Total} & \colhead{\scriptsize Inner} & \colhead{\scriptsize Outer\ \ \ \ } & \colhead{\scriptsize \ \ \ \ Total} & \colhead{\scriptsize Inner} & \colhead{\scriptsize Outer\ \ \ \ } & \colhead{\scriptsize \ \ \ \ Total} & \colhead{\scriptsize Inner} & \colhead{\scriptsize Outer\ \ \ \ } & \colhead{\scriptsize \ \ \ \ Total} & \colhead{\scriptsize Inner} & \colhead{\scriptsize Outer\ \ \ \ } & \colhead{\scriptsize \ \ \ \ Total} & \colhead{\scriptsize Inner} & \colhead{\scriptsize Outer\ \ \ \ }\\
\colhead{(1)} & \colhead{\ \ \ \ (2)} & \colhead{(3)} & \colhead{(4)\ \ \ \ } & \colhead{\ \ \ \ (5)} & \colhead{(6)} & \colhead{(7)\ \ \ \ } & \colhead{\ \ \ \ (8)} & \colhead{(9)} & \colhead{(10)\ \ \ \ } & \colhead{\ \ \ \ (11)} & \colhead{(12)} & \colhead{(13)\ \ \ \ } & \colhead{\ \ \ \ (14)} & \colhead{(15)} & \colhead{(16)\ \ \ \ }
}
\startdata
IC~4710 & 0.69 & 0.74 & \nodata & $-$0.09 & $-$0.02 & \nodata & $-$0.16 & $-$0.41 & $-$0.77 & \nodata & \nodata & $-$0.14 & \nodata & \nodata & \nodata \\
Mrk~33 & $-$0.12 & $-$0.12 & $-$0.10 & $-$0.45 & $-$0.42 & $-$0.60 & 0.06 & 0.14 & 0.03 & $-$0.09 & $-$0.23 & 0.11 & $-$1.58 & $-$2.16 & $-$2.19 \\
NGC~24 & $-$0.37 & $-$0.71 & 0.02 & $-$0.30 & \nodata & \nodata & $-$0.17 & $-$0.09 & \nodata & \nodata & \nodata & \nodata & \nodata & \nodata & \nodata \\
NGC~337 & $-$0.37 & $-$0.42 & $-$0.39 & $-$0.71 & $-$0.49 & $-$0.33 & $-$0.20 & $-$0.08 & $-$0.09 & \nodata & 0.21 & \nodata & \nodata & $-$1.94 & \nodata \\
NGC~628 & $-$0.82 & \nodata & \nodata & 0.27 & \nodata & \nodata & $-$0.20 & $-$0.34 & 0.06 & \nodata & \nodata & \nodata & \nodata & \nodata & \nodata \\
NGC~855 & $-$0.74 & 0.11 & 0.12 & 0.31 & $-$0.62 & $-$0.61 & $-$0.06 & 0.06 & 0.03 & \nodata & \nodata & \nodata & \nodata & \nodata & \nodata \\
NGC~925 & $-$0.31 & $-$0.40 & $-$0.31 & $-$0.49 & \nodata & \nodata & $-$0.29 & $-$0.12 & $-$0.03 & \nodata & \nodata & \nodata & \nodata & \nodata & \nodata \\
NGC~1097 & $-$1.07 & $-$0.89 & $-$1.13 & $-$0.84 & $-$1.11 & $-$1.25 & $-$0.19 & 0.01 & $-$0.01 & 0.20 & 0.64 & 0.42 & $-$1.69 & $-$1.75 & $-$1.94 \\
NGC~1266 & $-$0.48 & $-$0.54 & $-$0.33 & $-$0.73 & \nodata & \nodata & $-$0.53 & \nodata & 0.04 & \nodata & \nodata & \nodata & \nodata & \nodata & \nodata \\
NGC~1291 & 0.14 & 0.09 & 0.18 & $-$0.63 & $-$0.58 & $-$0.44 & $-$0.06 & 0.08 & \nodata & 0.11 & 0.73 & \nodata & $-$1.09 & $-$1.13 & \nodata \\
\enddata
\tablecomments{\footnotesize For each line ratio, we give three measurements: integrated over the central region, the inner region (central $10\arcsec \times 10\arcsec$ of low-resolution datacube after projection), and the outer region (remaining area with complete high-resolution observations).  (This table is available in its entirety in machine-readable form.)}
\label{tab:FluxRatioI}
\end{deluxetable*}

\begin{figure*}[!ht]
\center{\includegraphics[width=1.0\textwidth]{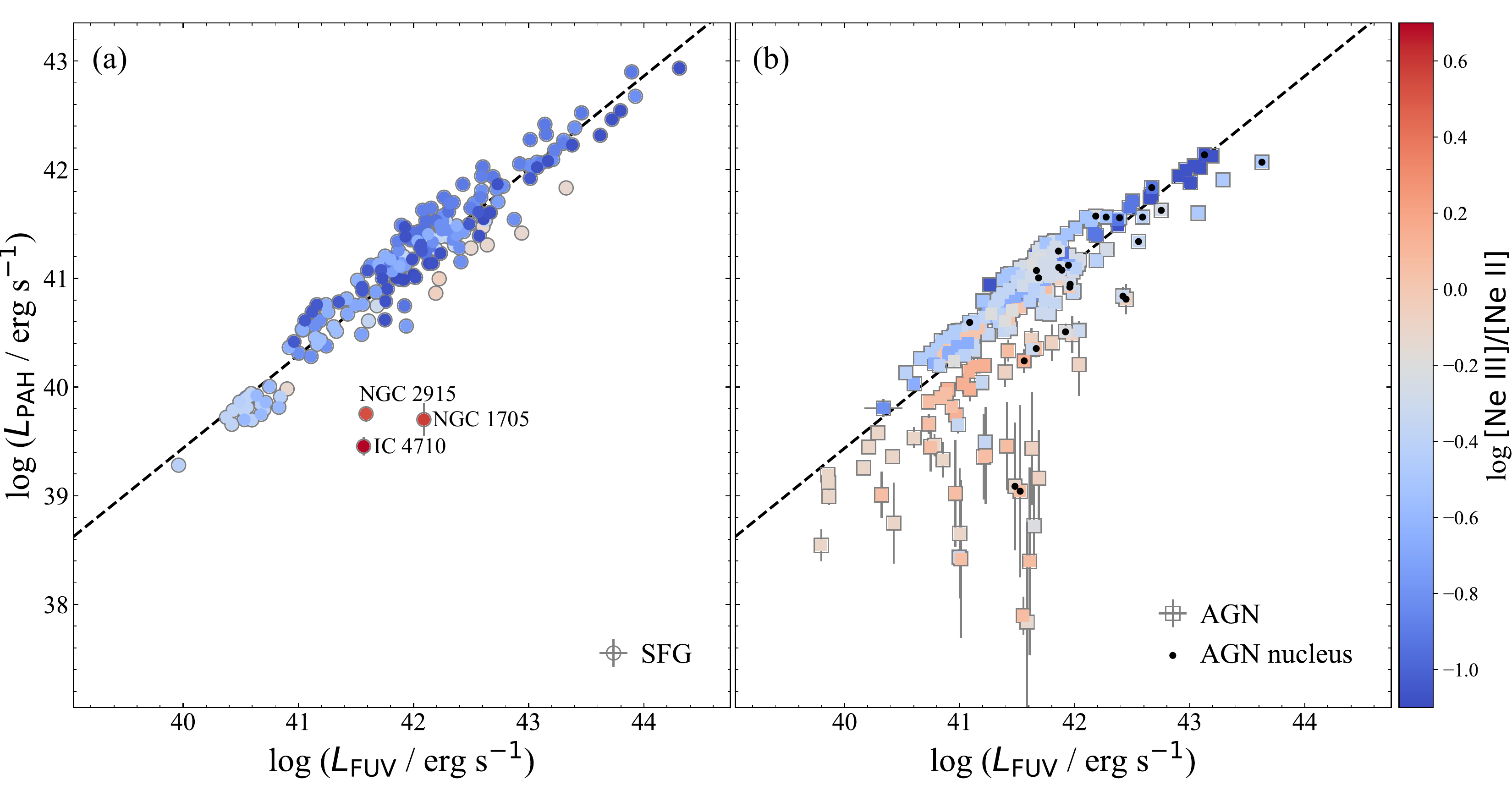}}
\caption{Correlation between integrated PAH luminosity and extinction-corrected FUV luminosity for spaxels in galaxies classified as (a) SFGs and (b) AGNs.  The data points are color-coded according to [Ne\,{\small III}]/[Ne\,{\small II}]. In panel (b), the points with a black dot pertain to the centralmost spaxel of the AGN. The dashed line is the best-fit linear regresson for the SFG spaxels (Equation~1).}
\label{fig:PAH_FUV_Ne2}
\end{figure*}

\begin{figure*}[!ht]
\center{\includegraphics[width=0.86\textwidth]{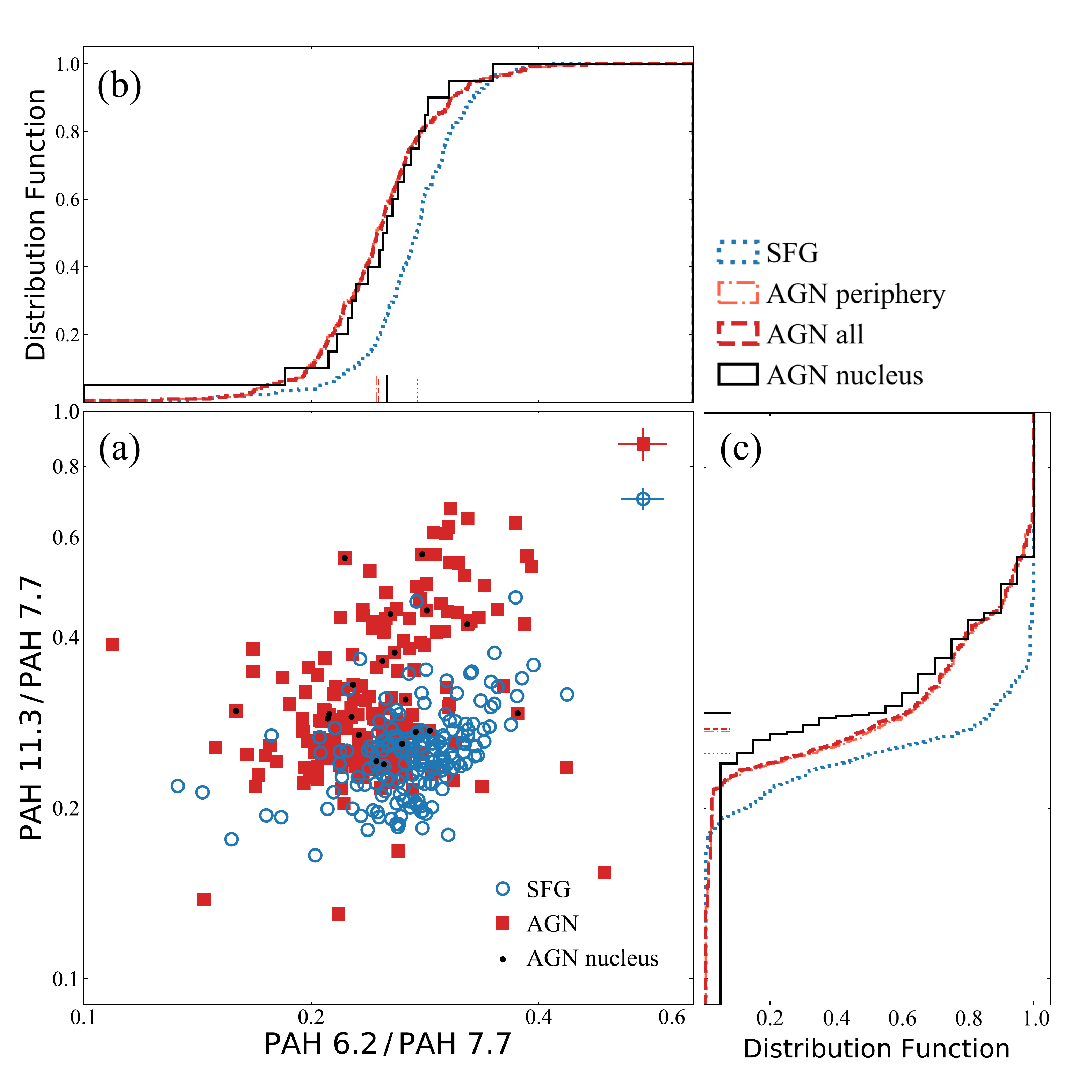}}
\caption{(a) Spatially resolved diagnostics of PAH band ratios 6.2\,$\mum$/7.7\,$\mum$ versus 11.3\,$\mum$/7.7\,$\mum$ for spaxels in galaxies classified as SFGs (blue) and AGNs (red). Points with a black dot pertain to the centralmost spaxel of each AGN (AGN nucleus).  AGN periphery are spaxels from AGNs that exclude the nucleus and cover the periphery of each AGN. Panels (b) and (c) respectively show the cumulative distribution of 6.2\,$\mum$/7.7\,$\mum$ and 11.3\,$\mum$/7.7\,$\mum$ band ratios, with the median of each distribution indicated by the corresponding short line.}
\label{fig:PAH_ratio_statistics}
\end{figure*}

\section{Spatially Resolved Analysis}\label{section:sec3}

We use the spatially resolved measurements to investigate the impact of AGN activity on PAH emission.  Among the 453 spaxels resolved within the sample of 66 galaxies, 269 are extracted from the central regions of galaxies classified as AGNs, while the remaining 184 derive from the central regions of SFGs. Although we refer to these as AGN spaxels and SFG spaxels, respectively, we stress that not all the AGN spaxels are necessarily strictly powered by the central active nucleus.  Apart from the centralmost spaxel on which the published nuclear classification was based, the current lack of optical integral-field spectroscopic observations of the same regions mapped by the IRS prevents us from ascertaining the dominant ionization mechanism of the non-nuclear spaxels.  While the radiation field of the AGN can reach significant physical extension depending on the power of the ionizing source (e.g., \citealt{Chen et al. 2019, Molina et al. 2022}), the ionization is highly anisotropic, and some of the peripheral spaxels could be powered by other sources of ionization.

\subsection{Evidence that AGNs Suppress PAH Emission}\label{sec:section3.1}

Previous work, based on integrated spectra, has revealed that PAH emission is relatively weaker in AGN environments, plausibly due to the destruction of small grains by their hard radiation field (e.g., \citealt{Smith et al. 2007b}; \citealt{ODowd et al. 2009}).  We aim to shed light on this issue using our spatially resolved analysis.  We begin by establishing the baseline level of PAH emission associated with the star-forming spaxels in inactive galaxies. Figure~\ref{fig:PAH_FUV_Ne2}a confirms, as already well-established using integrated spectra (e.g., \citealt{Shipley et al. 2016, Xie & Ho 2019}) and more limited explorations of spatially resolved spectra (e.g., \citealt{Maragkoudakis et al. 2018, Zhang et al. 2021}), that PAH emission correlates strongly with star formation activity.  We use the extinction-corrected FUV emission to gauge the level of ongoing star formation, avoiding H$\alpha$, even though it is available for the SINGS sample, to obviate contamination by nebular emission powered by AGNs in our ensuing comparative analysis.  To account for Galactic extinction, we adopt the extinction curve of \cite{Cardelli et al. 1989} with $R_V = 3.1$, and $E(B-V)$ from \citealt{Schlafly & Finkbeiner 2011}\footnote{\url{https://irsa.ipac.caltech.edu/applications/DUST/}} (\citealt{IRSA 2022}), and $A_{\rm FUV} = 7.9\ E(B-V)$ \citep{Gil de Paz et al. 2007}.  Correction for internal extinction follows $L_{\rm FUV} = L_{\rm FUV}^{\rm obs} + 6.0\,L_{\rm 24\,\mum}$ \citep{Liu et al. 2011}, where $L_{\rm FUV}^{\rm obs}$ has already been corrected for Galactic extinction.  The points are color-coded by the [Ne\,{\small III}]/[Ne\,{\small II}] line ratio, a measure of the hardness of the radiation field. All else being equal, a harder radiation field, which supplies relatively more high-energy photons, produces higher [Ne\,{\small III}]/[Ne\,{\small II}] (\citealt{Thornley et al. 2000}).  We note that spaxels with higher [Ne\,{\small III}]/[Ne\,{\small II}] tend to have systematically lower $L_{\rm PAH}$ at fixed $L_{\rm FUV}$, with the highest excitation and largest PAH deficit seen in three late-type, dwarf galaxies (IC\,4710, NGC\,1705, and NGC\,2915), as expected from the inverse correlation between nebular excitation and metal abundance (e.g., \citealt{OHalloran et al. 2006, Xie & Ho 2019}).  A hard radiation field can destroy PAH molecules \citep{Gordon et al. 2008, Hunt et al. 2010, Li & Draine 2002}. 

A linear regression of the SFG spaxels with the {\tt Python} package {\tt linmix} (\citealt{Kelly 2007}) yields the following relation, which has an intrinsic scatter of $\rm 0.24\ dex$:

\begin{align}\label{equ:calHa}
\begin{aligned}
&{\rm log}\,L_{\rm PAH} = \\&(0.856\pm0.022)({\rm log}\, L_{\rm FUV} - 41.5) + (40.722\pm0.021).
\end{aligned}
\end{align}

\noindent
Consistent with the spatially resolved analysis of M\,51 \citep{Zhang et al. 2021}, the best-fitting slope is less than 1. Such a sub-linear slope may be a consequence of PAH excitation by evolved stars in regions of low star formation activity \citep{Zhang & Ho 2022}. 

The spaxels from the central regions of AGNs show a markedly different behavior (Figures~\ref{fig:PAH_FUV_Ne2}b).  While the majority of the spaxels occupy a distribution whose upper envelope closely follows the best-fit $L_{\rm PAH}-L_{\rm FUV}$ relation of SFGs, a subset of points deviates markedly below the best-fit line, by as much as $\Delta \log\, L_{\rm PAH} \approx 2-3$~dex.  The suppression of PAH emission is generally associated with spaxels of relatively high excitation ($\rm \log~[Ne\,{\small III}]/[Ne\,{\small II}] \simeq 0$), but, interestingly, the degree of PAH suppression in AGNs can be much more extreme than in star-forming dwarf galaxies, even though the latter can have higher values of excitation than the former.  Moreover, among the deviant AGN spaxels, we see no systematic dependence between the degree of PAH deficit and [Ne\,{\small III}]/[Ne\,{\small II}].  These results suggest that some factor other than the hardness of the radiation field---one that is uniquely associated with black hole accretion---affects the PAH properties in AGNs.

\subsection{Evidence that AGNs Affect the PAH Band Ratios}\label{sec:section3.2}

Theoretical calculations show that the size distribution and ionization state of PAH molecules play an important role in determining the relative strength of their spectral features \citep{Draine & Li 2001, Draine & Li 2007}. Smaller PAHs, upon absorbing a UV photon, reach much higher levels of vibrational excitation and radiate at shorter, more energetic wavelengths because of their low heat capacity. On the other hand, PAHs with a larger neutral fraction generate higher values of the 11.3\,$\mum$/7.7\,$\mum$ ratio, as neutral PAHs produce stronger C--H modes responsible for the 3.3\,$\mum$ and 11.3\,$\mum$ PAH features, while the C--C modes of cationic PAHs efficiently emit the $6 - 9\ \mum$ PAH features \citep{Draine & Li 2007, Tielens 2008}. PAH spectra also depend on the age of the stellar population responsible for their excitation, with the 6.2\,$\mum$/7.7\,$\mum$ ratio dropping by a factor $\sim 1.5$ as the population varies from a 3 Myr starburst to the advanced evolutionary state of M\,31's bulge (\citealt{Draine et al. 2021}).

While to first order SFGs and AGNs share similar PAH spectral characteristics, on closer inspection it is apparent that the spaxels of the two classes of objects have statistically different distributions of PAH band ratios (Figure~\ref{fig:PAH_ratio_statistics}).  Relative to the spaxels in SFGs, AGN spaxels tend to have somewhat lower 6.2\,$\mum$/7.7\,$\mum$ ratios and significantly stronger 11.3\,$\mum$/7.7\,$\mum$ ratios. To quantify the statistical difference between the band ratios of the two groups, we examine their cumulative distributions in Figures~\ref{fig:PAH_ratio_statistics}b and \ref{fig:PAH_ratio_statistics}c.  Based on the Kolmogorov-Smirnov test with $p < 10^{-6}$, we can reject the null hypothesis that the two samples are drawn from the same parent distribution. The stronger 11.3\,$\mum$/7.7\,$\mum$ ratio is even more pronounced for the centralmost spaxel (AGN nucleus) after we distinguish it from its peripheral spaxels, although the lower 6.2\,$\mum$/7.7\,$\mum$ becomes less pronounced. The differences between the band ratios of the two groups support the notion that the spectral peculiarities of active galaxies are driven by factors related to the nuclear region, as further explored in the following section.

\begin{figure*}[!ht]
\center{\includegraphics[width=1\textwidth]{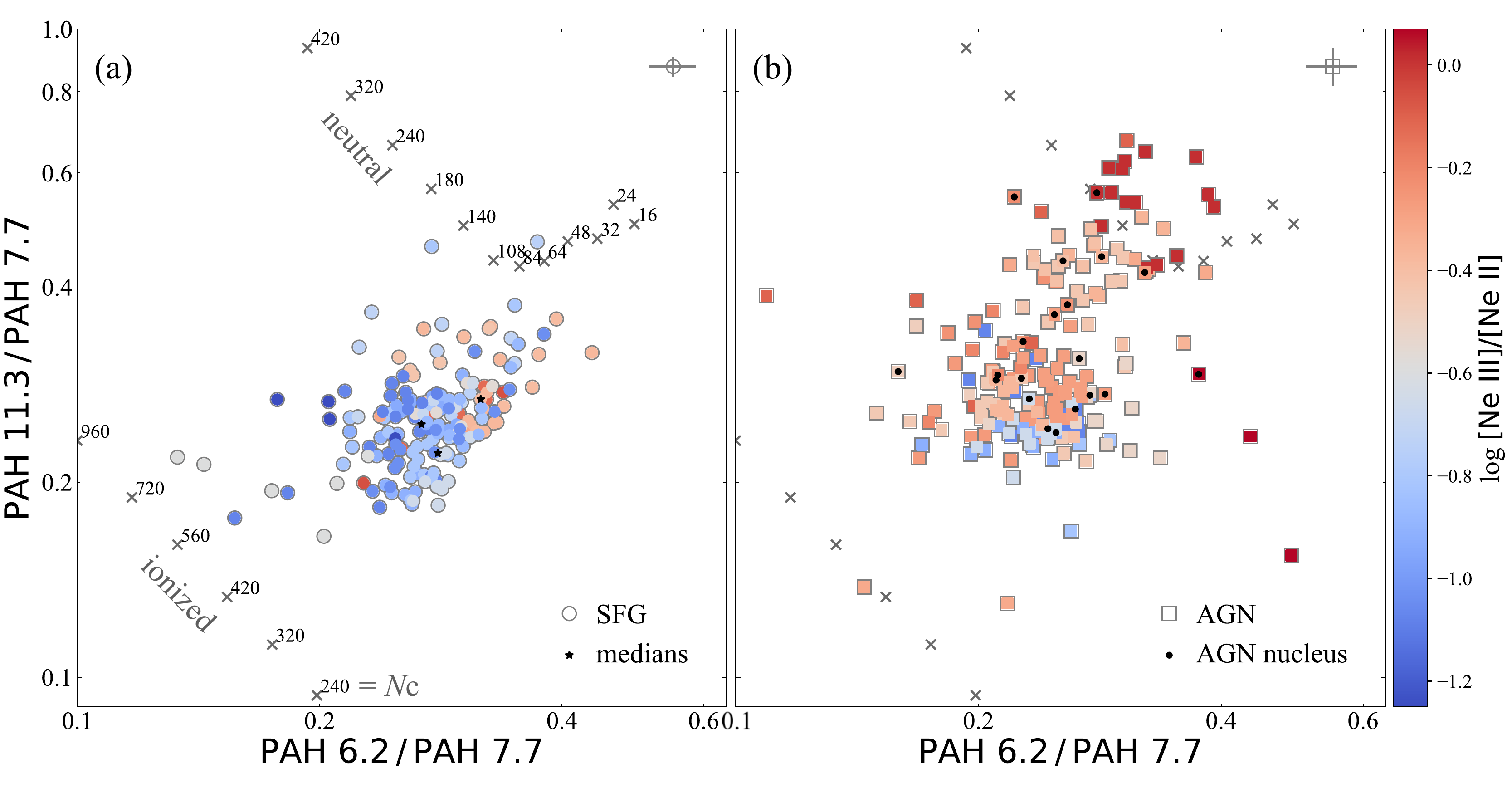}}
\caption{Spatially resolved diagnostics of PAH band ratios 6.2\,$\mum$/7.7\,$\mum$ and 11.3\,$\mum$/7.7\,$\mum$ for spaxels in galaxies classified as (a) SFGs and (b) AGNs, color-coded according to the [Ne\,{\small III}]/[Ne\,{\small II}] line ratios. In panel (a), the three black stars, from left to right, give the median PAH band ratios for SFG spaxels that cover the range $\rm log\ [Ne\,{\small III}]/[Ne\,{\small II}]<-0.7$, $\rm -0.7 < log\ [Ne\,{\small III}]/[Ne\,{\small II}] < -0.5$, and $\rm log\ [Ne\,{\small III}]/[Ne\,{\small II}]>-0.5$. In panel (b), the points with a black dot pertain to the centralmost spaxel of the AGN. The black crosses are model predictions by \cite{Draine & Li 2001} for neutral (top sequence) and ionized (bottom sequence) PAHs, as well as PAHs of different sizes, with increasing number of C atoms ($N_{\rm C}$) from right to left. Typical error bars of the PAH band ratios are marked in the top-right of each panel.
}\label{fig:PAH_Ratio_CenS}
\end{figure*}

\section{Discussion}\label{section:sec4}

A central aim of this work is to investigate the extent to which the properties of PAH emission in active galaxies differ from those in normal, star-forming galaxies, with the hope of elucidating the factors that affect the detailed mechanisms that govern PAH emission in extragalactic environments, which, in turn, may provide us with new diagnostic tools to probe physical processes in galaxy evolution.  Our systematic comparison of nearby galaxies with mapping-mode IRS observations has revealed two clear trends between galaxies with and without an active nucleus: the presence of an AGN tends to suppress the overall level of PAH emission (Section~\ref{sec:section3.1}) and produce subtle differences in the relative intensities of the main PAH features (Section~\ref{sec:section3.2}).  Here we explore the possible physical drivers of these observed trends.  

\subsection{Influence on PAH Band Ratios from Radiative Effects}

We begin by establishing, as a baseline reference, the behavior of the PAH band ratios 6.2\,$\mum$/7.7\,$\mum$ and 11.3\,$\mum$/7.7\,$\mum$ in galaxies classified as SFGs, in response to variations in nebular excitation as judged by the [Ne\,{\small III}]/[Ne\,{\small II}] ratio (Figure~\ref{fig:PAH_Ratio_CenS}a)\footnote{In this and subsequent sections, we omit a few objects whose individual PAH features are too uncertain ($\gtrsim 0.5$~dex) to yield reliable band ratios. In a few extreme cases, the flux of an individual PAH feature may even formally exceed the integrated flux of all PAH emission derived from the template decomposition. The spectra of the SFG NGC\,1377 are too difficult to measure because of strong silicon absorption (\citealt{Roussel et al. 2006}).  We exclude all the spaxels of the AGNs NGC\,1291, 1316, 1404, 4125, 4552, 4594, and 4725, and some spaxels of IC\, 4710, Mrk\,33, NGC\,3031 (i.e., spaxels with only integrated PAH luminosity as shown in Table~\ref{tab:TablePAH}), because their spectra are so starlight-dominated or so noisy that it proved impossible to measure robust PAH band ratios. The omitted spaxels account for 12\% of the sample spaxels.}. A higher value of [Ne\,{\small III}]/[Ne\,{\small II}] signifies a harder radiation field, one whose SED has relatively more high-energy photons. We overlay theoretical tracks from the model predictions of \cite{Draine & Li 2001} for neutral (top sequence) and ionized (bottom sequence) PAHs of different size, as parameterized by the number of C atoms, $N_{\rm C}$.  It is immediately apparent that the spatially resolved regions within SFGs have rather uniform spectra, as evidenced by the fact that the majority of the spaxels are clustered around a small range of values for the two band ratios\footnote{We excluded the spaxels from the dwarf galaxies to provide a clearer view of the color bar.}. Moreover, subtle dependence on nebular excitation can be seen.  The 6.2\,$\mum$/7.7\,$\mum$ ratio mildly increases with increasing [Ne\,{\small III}]/[Ne\,{\small II}]. The behavior of 11.3\,$\mum$/7.7\,$\mum$ is more complex: it first appears to decrease with [Ne\,{\small III}]/[Ne\,{\small II}], and then the trend reverses, such that the blueish and reddish points are located on the top of the diagram, while the grayish points with $\rm log\ [Ne\,{\small III}]/[Ne\,{\small II}] \approx -0.6$ cluster more to the bottom. To help visualize this complex behavior, we divide the SFG spaxels into three subgroups according to their neon line ratios, and then plot the median of the PAH band ratios for the three groups as the black stars in Figure~\ref{fig:PAH_Ratio_CenS}a.

How does excitation affect 6.2\,$\mum$/7.7\,$\mum$?  A harder illuminating spectrum contains more UV photons at higher energy and hence directly enhances PAH emission at shorter wavelengths relative to that at longer wavelengths \citep{Draine et al. 2021, Rigopoulou et al. 2021}. If this were true, the 11.3\,$\mum$/7.7\,$\mum$ band ratio should decrease strictly with increasing excitation, because a harder radiation field preferentially enhances PAH emission at shorter wavelengths and would boost the fraction of ionized PAH molecules. However, as explained above, the variation of 11.3\,$\mum$/7.7\,$\mum$ with [Ne\,{\small III}]/[Ne\,{\small II}] is not monotonic. Alternatively, a harder radiation field can modify PAH size distribution through preferential photo-erosion of smaller PAH grains (\citealt{Allain et al. 1996a, Micelotta et al. 2010a, Micelotta et al. 2010b, Murga et al. 2016}). The normally narrow peak size distribution of 150 C atoms for astronomical PAHs \citep{Weingartner & Draine 2001a, Draine & Li 2007} would shift with photo-erosion toward sizes of fewer than 100 C atoms (with total amount of smaller PAHs reduced as well). This effect also can potentially account for the increase of 6.2\,$\mum$/7.7\,$\mum$, but it cannot explain why 11.3\,$\mum$/7.7\,$\mum$ does not simultaneously increases in lockstep. The 11.3\,$\mum$/7.7\,$\mum$ band ratio first decreases and {\it then}\ increases with [Ne\,{\small III}]/[Ne\,{\small II}]. To account for this peculiar behavior, we propose that, during the photo-erosion process, a harder radiation field first elevates the ionization fraction of the PAH molecules up to a certain threshold, which results in the decrease of 11.3\,$\mum$/7.7\,$\mum$, beyond which photo-erosion proceeds to dominate by modifying the PAH size distribution, which results in the increase of 6.2\,$\mum$/7.7\,$\mum$ and 11.3\,$\mum$/7.7\,$\mum$. In addition, preferential photo-destruction of the ionized PAH grains, which are more vulnerable relative to the neutral ones (\citealt{Allain et al. 1996b, Holm et al. 2011}), also may contribute to the increase of 11.3\,$\mum$/7.7\,$\mum$. This scenario further provides a plausible explanation for the severe depletion of the overall PAH emission among the dwarf galaxies with the hardest radiation field (Figure~\ref{fig:PAH_FUV_Ne2}a).

By sharp contrast, AGN spaxels occupy a much broader distribution of PAH band ratios (Figure~\ref{fig:PAH_Ratio_CenS}b), displaying characteristically larger 11.3\,$\mum$/7.7\,$\mum$ but smaller 6.2\,$\mum$/7.7\,$\mum$ compared to the SFG population.  No clear correlation exists with [Ne\,{\small III}]/[Ne\,{\small II}], except for the slight tendency for the highest 11.3\,$\mum$/7.7\,$\mum$ ratios to be associated with the most enhanced [Ne\,{\small III}]/[Ne\,{\small II}], most notably for the data points that exceed the upper boundary of model predictions by \cite{Draine & Li 2001}.  Given the same range of [Ne\,{\small III}]/[Ne\,{\small II}], the distributions of PAH band ratios are significantly different for the AGN spaxels compared to the SFG spaxels.  While the overall population shift toward lower 6.2\,$\mum$/7.7\,$\mum$ and higher 11.3\,$\mum$/7.7\,$\mum$ can be broadly understood in terms of a grain population biased in favor of larger species \citep{Draine & Li 2007, Draine et al. 2021}, the absence of clear trends with excitation suggests that radiative effects alone cannot fully explain the PAH characteristics of the AGN spaxels. Similarly, although the relative fraction of ionized versus neutral PAH molecules can affect the band ratios, we also regard variation in ionization fraction to be an unlikely culprit for the observed spectral differences between active and inactive galaxies, for the simple fact that the harsher environment anticipated in the vicinity of an AGN should produce more ionized PAHs, and hence weaker---not stronger---11.3\,$\mum$/7.7\,$\mum$. We expect AGN spaxels to gravitate toward the theoretical sequence for ionized PAHs instead of the one for neutral PAHs (bottom instead of top of Figure~\ref{fig:PAH_Ratio_CenS}b).

\begin{figure}[!t]
\center{\includegraphics[width=1\linewidth]{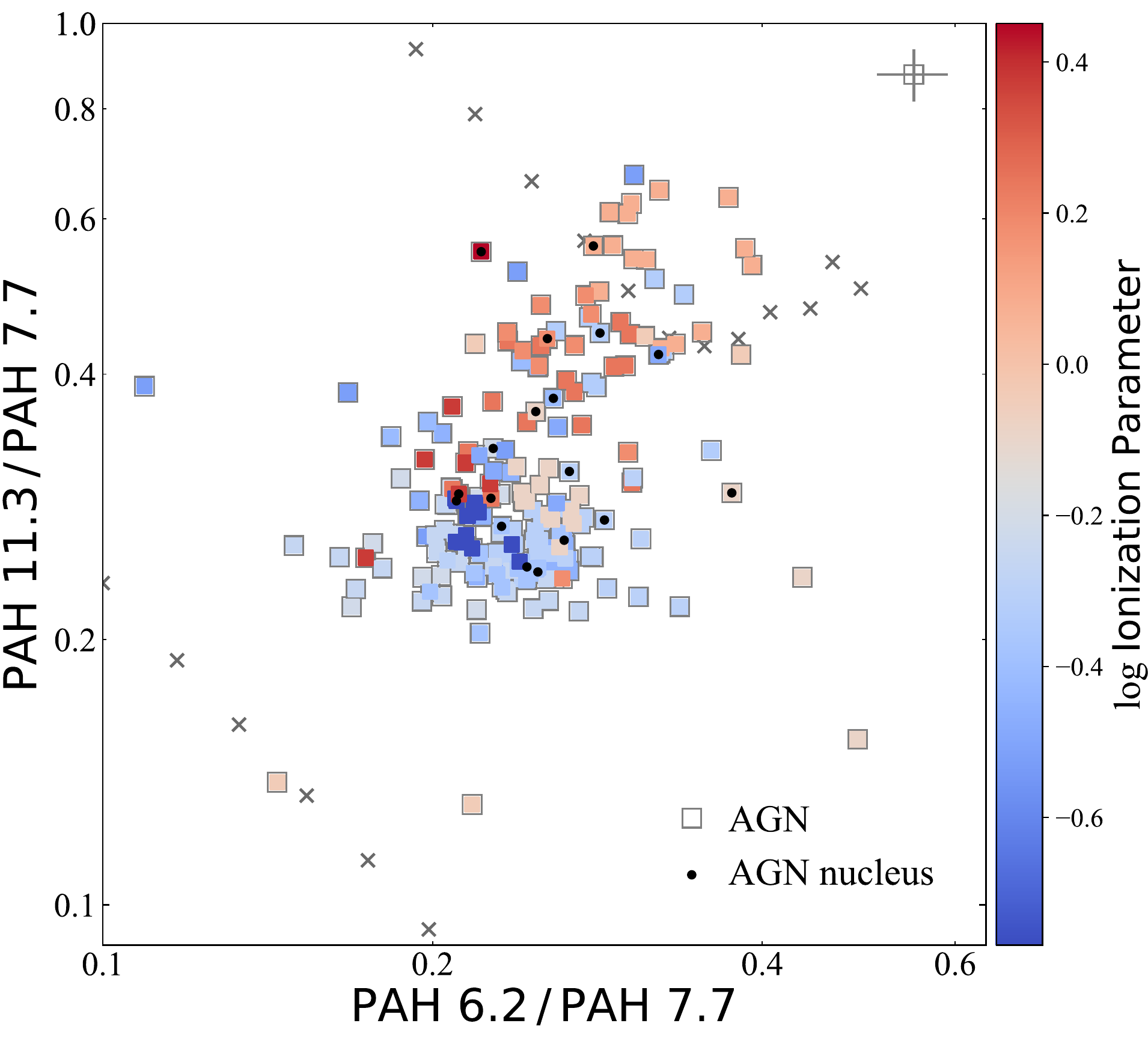}}
\caption{{The PAH band ratios 6.2\,$\mum$/7.7\,$\mum$ and 11.3\,$\mum$/7.7\,$\mum$ for AGN spaxels that have measurements of [S\,{\small IV}]/[Ne\,{\small III}], H$_{2}\,S(2)$/H$_{2}\,S(0)$, and [S\,{\small III}]\,18.71\,$\mum$/[S\,{\small III}]\,33.48\,$\mum$, color-coded according to the ionization parameter calculated based on these line ratios.}\label{fig:PAH_Ratio_AGN_IonBalan}}
\end{figure}

\begin{figure*}
\center{\includegraphics[width=1\textwidth]{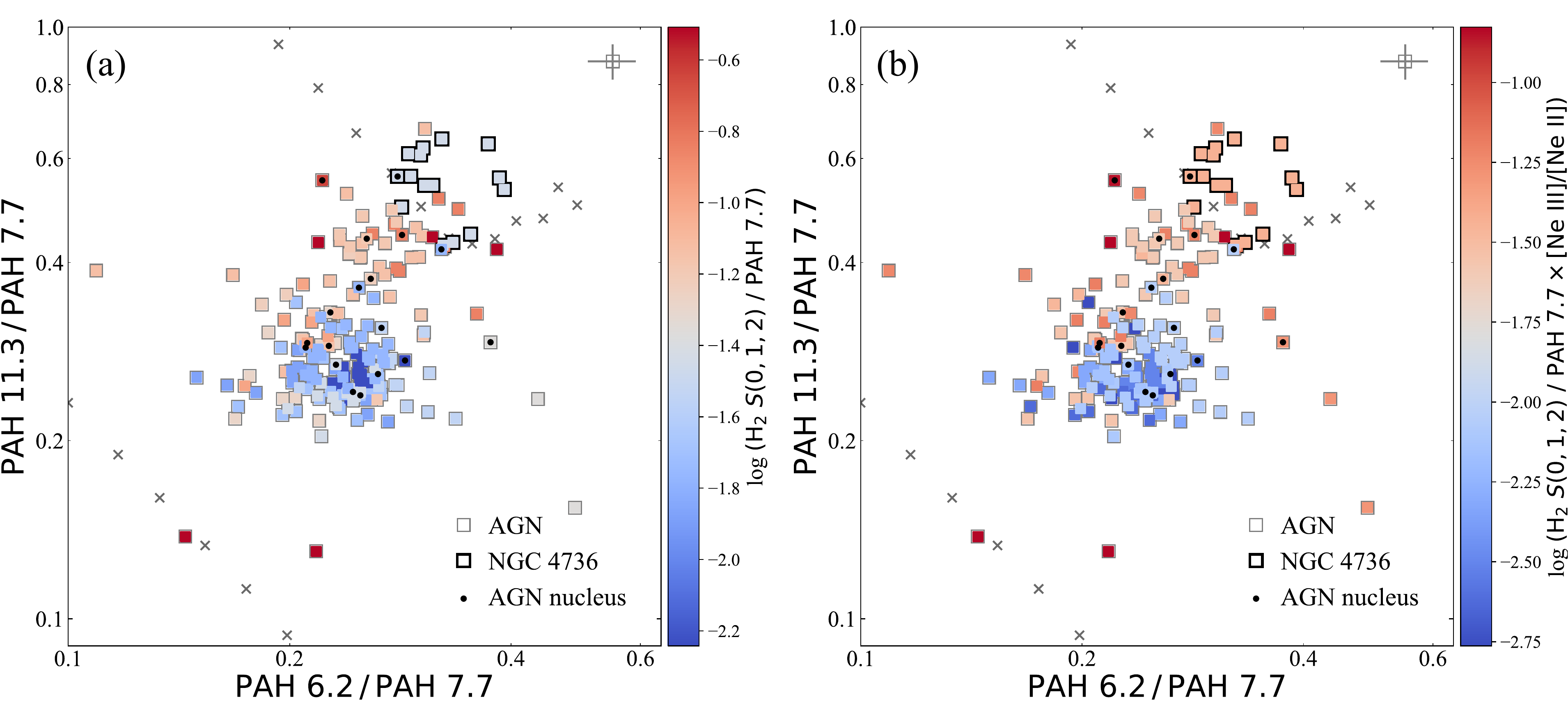}}
\caption{The PAH band ratios 6.2\,$\mum$/7.7\,$\mum$ and 11.3\,$\mum$/7.7\,$\mum$ for AGN spaxels, color-coded according to (a) the line ratio H$_{2}\,S(0,1,2)$ / PAH\,7.7\,$\mum$, which is an indicator of shock strength, and (b) the combination of H$_{2}\,S(0,1,2)$ / PAH\,7.7\,$\mum \times \rm [Ne\,{\small III}]/[Ne\,{\small II}]$, which is an indicator of the combined effects of shocks and radiative effects.
}\label{fig:PAH_Ratio_AGN_shocks}
\end{figure*}

\begin{figure*}[!t]
\center{\includegraphics[width=1\textwidth]{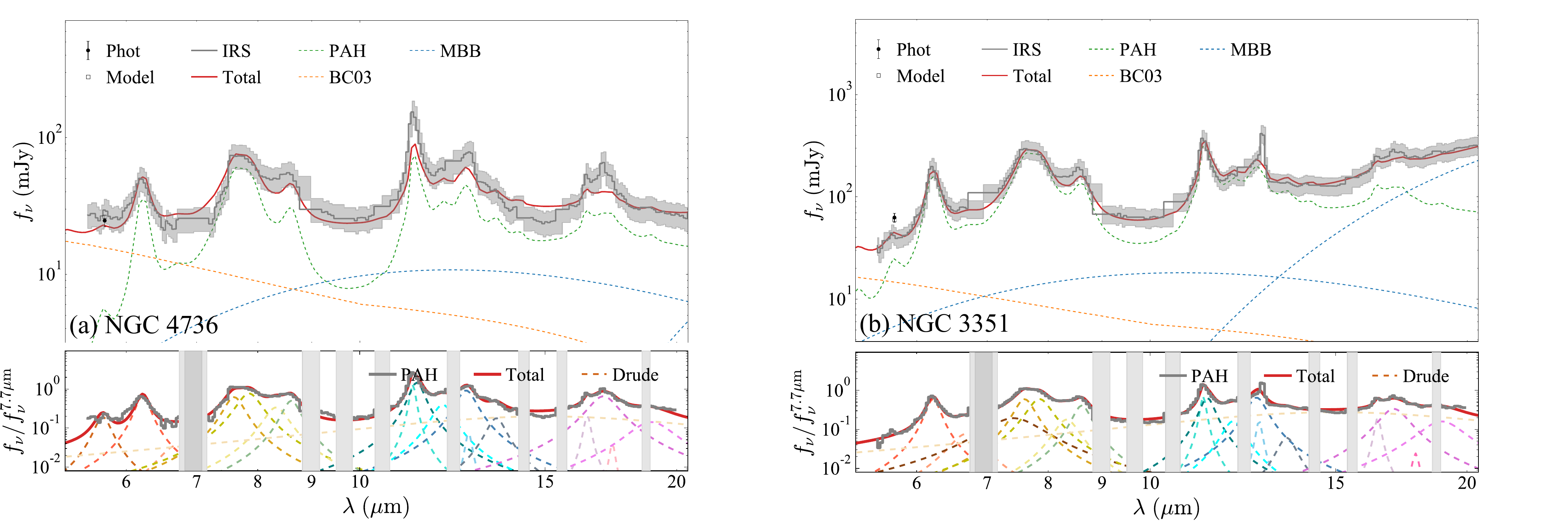}}
\caption{Illustration of the PAH decomposition for (a) NGC\,4736 and (b) NGC\,3351. NGC 4736 exhibits stronger PAH emission at 11.3 $\mum$ and longer wavelengths than NGC 3351.}
\label{fig:PAHComp}
\end{figure*}

Nevertheless, we examine whether ionization effects can produce the stronger 11.3\,$\mum$/7.7\,$\mum$ ratios for the AGN spaxels. The ionization balance is conventionally described by $G_{0} T^{1/2}/n_{e}$, where $G_{0}$ is the intensity of the radiation field, $T$ is the gas temperature, and $n_{e}$ is the electron density \citep{Bakes & Tielens 1994, Weingartner & Draine 2001b}.  We compute the modified quantity $G_{0}^{*}T^{*}/n_{e}^{*}$ as an observational proxy for the ionization parameter.\footnote{Parameters denoted by * serve as an empirical surrogate for the actual physical parameters.} We represent $G_{0}^{*}$ using [S\,{\small IV}]\,10.5\,$\mum$/[Ne\,{\small III}]\,15.5\,$\mum$, a line ratio that tightly correlates with the dimensionless ionization parameter \citep{Pereira-Santaella et al. 2017}. In light of the scaling relation between gas temperature and the theoretically calculated intensity ratio of the high-energy to low-energy rotation-vibration transitions of molecular hydrogen \citep{Turner et al. 1977, Vega et al. 2010}, we assume that $T^{*}$ is proportional to H$_{2}\,S(2)$\,12.3\,$\mum$/H$_{2}\,S(0)$\,28.2\,$\mum$.  Finally, we use [S\,{\small III}]\,18.7\,$\mum$/[S\,{\small III}]\,33.5\,$\mum$ to approximate $n_{e}^{*}$, as these two fine-structure lines are density-sensitive because they come from the same ion but have different critical densities \citep{Dale et al. 2006, Pereira-Santaella et al. 2017}.  Figure~\ref{fig:PAH_Ratio_AGN_IonBalan} shows the distribution of PAH band ratios color-coded according to the modified ionization parameter. This result strongly conflicts with the actual observations, which show that the regions with high ionization parameter actually exhibit large 11.3\,$\mum$/7.7\,$\mum$ ratios, more consistent with neutral instead of ionized PAHs. 

\subsection{Influence on PAH Band Ratios from Mechanical Effects}\label{sec:section4.2}

Having failed to establish a clear link between the PAH characteristics of AGN spaxels and radiative effects, we turn our attention to the possible role of mechanical effects from shocks, which can effectively destroy smaller PAH molecules and lead to strong variations in the strength of the $5-8\ \mum$ PAH features in AGNs \citep{Diamond-Stanic & Rieke 2010, Micelotta et al. 2010a, Micelotta et al. 2010b}. \cite{Roussel et al. 2007} found that the intensities of the IR H$_{2}$ lines scale tightly with PAH emission over a large range of radiation field intensity, conceivably because both tracers originate predominantly in either dense or diffuse photo-dissociation regions (PDRs).  Moreover, Roussel et al. proposed that shock heating contributes to an excess of IR molecular hydrogen emission. Figure~\ref{fig:PAH_Ratio_AGN_shocks}a showcases the PAH band ratios of the spaxels extracted from active galaxies whose central region has reliable measurements of the H$_{2}\,S(0)$, H$_{2}\,S(1)$, and H$_{2}\,S(2)$ emission lines, where now the color encodes the diagnostic ratio H$_{2}\,S(0,1,2)$/PAH\,7.7 used to indicate shock strength. Here, H$_{2}\,S(0,1,2)$ denotes the sum of all three molecular hydrogen vibrational lines. A distinctly bimodal distribution can be seen. For the AGN spaxels with 11.3\,$\mum$/7.7\,$\mum$ $\lesssim 0.3$ (as well as the SFG spaxels not shown here), the H$_{2}\,S(0,1,2)$/PAH\,7.7 ratios are consistent with the theoretical values expected from PDRs (\citealt{Guillard et al. 2012, Stierwalt et al. 2014}). However, for most of the AGN spaxels with larger 11.3\,$\mum$/7.7\,$\mum$, their H$_{2}\,S(0,1,2)$/PAH\,7.7 ratios greatly exceed the threshold found in PDRs, $\rm log (H_{2}/PAH) = -1.4$.  We propose that the elevated H$_{2}\,S(0,1,2)$/PAH\,7.7 ratios originate from shocks, and that the direct destruction of small PAHs by shocks results in the larger 11.3\,$\mum$/7.7\,$\mum$ and smaller 6.2\,$\mum$/7.7\,$\mum$ ratios observed in LLAGNs.

NGC\,4736, a type~2 LINER \citep{Ho et al. 1997a} notable for its UV-bright nucleus \citep{Maoz et al. 1995} and recent burst of star formation \citep{Taniguchi et al. 1996}, stands out as a highly instructive outlier on account of its exceptionally strong 11.3\,$\mum$/7.7\,$\mum$ ratio and yet relatively {\it weak}\ shock strength. At the same time, this galaxy has the hardest radiation field among the AGN regions shown in Figure~\ref{fig:PAH_Ratio_CenS}b. Figure~\ref{fig:PAHComp} highlights the unique PAH characteristics of NGC\,4736, in contrast to NGC\,3351, which has the lowest [Ne\,{\small III}]/[Ne\,{\small II}] ratio in the sample, one that is more than 10 times lower than in NGC\,4736. This example illustrates that when mechanical effects are sub-dominant, a hard radiation field, whether it originates from star formation or AGN accretion, produces large 11.3\,$\mum$/7.7\,$\mum$ and 6.2\,$\mum$/7.7\,$\mum$ by the selective photo-erosion and photo-destruction of PAH molecules (see Section~4.1).  

We surmise that in general both radiative {\it and}\ mechanical effects operate to establish the spectral diversity of PAH emission in AGNs. The relative proportion of the two mechanisms, which depends on the specific attributes of individual systems and the intrinsic diversity in the composition of any galaxy sample, then determines the observed scatter. This hypothesis is strongly supported in Figure~\ref{fig:PAH_Ratio_AGN_shocks}b, where now the color coding represents a new quantity that combines the joint effects of radiation field hardness and shocks. A striking result emerges: the bimodal distribution of the PAH band ratios becomes even more distinctive than either effect alone.  In agreement with previous studies (e.g., \citealt{Diamond-Stanic & Rieke 2010, Micelotta et al. 2010a, Micelotta et al. 2010b}), we find that small grains, which emit more strongly at shorter wavelengths, are the most vulnerable to destruction by shocks.  Our current work reveals that radiative effects may serve an important supporting role, by contributing to the ionization and erosion of the PAH molecules and thereby making them even more vulnerable to destruction by shocks.

\begin{figure}[!t]
\center{\includegraphics[width=1\linewidth]{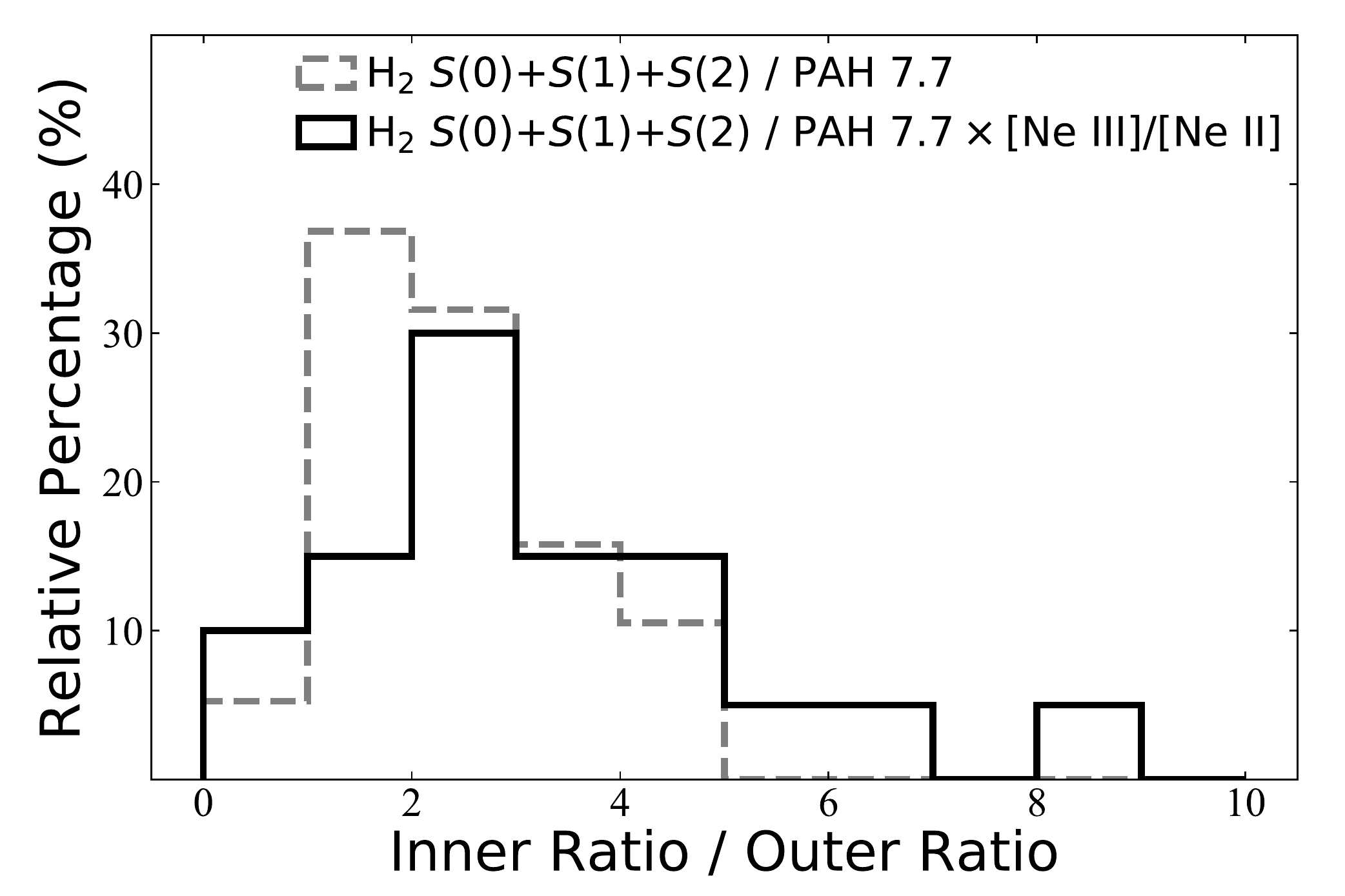}}
\caption{Statistics of mid-IR diagnostic line ratios as measured in the centralmost, inner ($10\arcsec \times 10\arcsec$) region compared to those measured in the remaining outer region. The line ratio H$_{2}\,S(0,1,2)$/PAH\,7.7\,$\mum$ is sensitive to shocks, while the combined diagnostic H$_{2}\,S(0,1,2)$/PAH\,7.7\,$\mum \times \rm [Ne\,{\small III}]/[Ne\,{\small II}]$ reflects the joint influence of shocks and radiative effects.  This plot illustrates that these indicators are even more prominent in the inner region.}
\label{fig:R_nuc_R_peri}
\end{figure}

\begin{figure*}
\center{\includegraphics[width=1\textwidth]{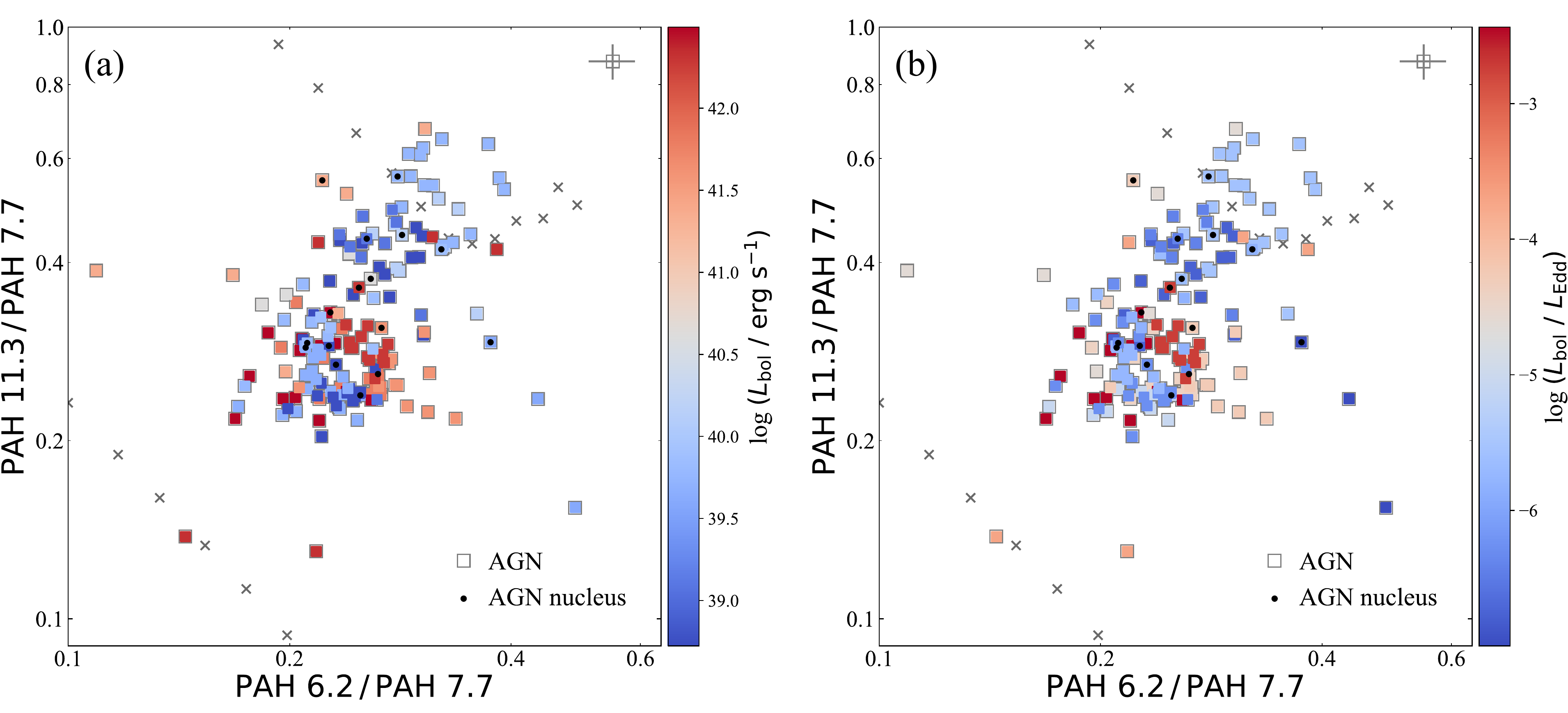}}
\caption{The PAH band ratios 6.2\,$\mum$/7.7\,$\mum$ and 11.3\,$\mum$/7.7\,$\mum$ for AGN spaxels that have measurements of the strength of nuclear activity, color-coded according to the (a) bolometric luminosity ($L_{\rm bol}$) and (b) Eddington ratio ($L_{\rm bol}/L_{\rm Edd}$) of the AGN.
}\label{fig:PAH_Ratio_AGN_AGNproperties}
\end{figure*}

\begin{figure*}
\center{\includegraphics[width=1\textwidth]{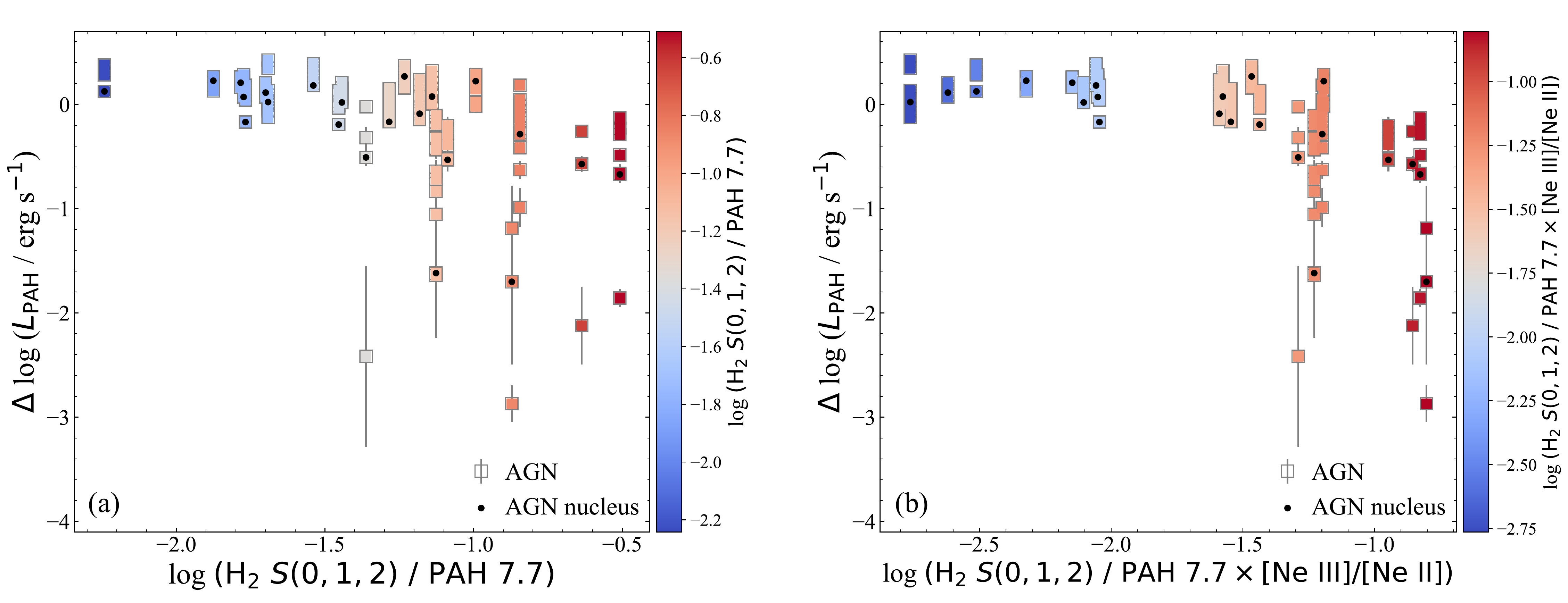}}
\caption{Dependence of PAH deficit of AGN spaxels, defined as the offset between the observed integrated ($5-20\ \mum$) PAH luminosity and that predicted from the extinction-corrected FUV luminosity based on the $L_{\rm PAH}-L_{\rm FUV}$ relation of SFG spaxels (Figure~\ref{fig:PAH_FUV_Ne2}; Equation~1) versus (a) shock strength, H$_{2}\ S(0,1,2)$/PAH\ 7.7\ $\mum$, and (b) the combined effects of shocks and radiative excitation, H$_{2}\ S(0,1,2)$/PAH\ 7.7\ $\mum \times \rm [Ne\,{\small III}]/[Ne\,{\small II}]$. The points with a black dot pertain to the centralmost spaxel of each AGN.}
\label{fig:PAH_deficit}
\end{figure*}

We reiterate an important caveat that applies to all the results discussed throughout this section. The physical diagnostics (i.e., for excitation, ionization parameter, and shocks) deduced from the high-resolution IRS observations are available only for a portion of the total area for which PAH emission was mapped in low-resolution mode.  We assume that the physical diagnostics extracted from the central region of each galaxy applies to all the spaxels spanning a larger area.  While we cannot verify in detail whether this assumption is correct, we can perform a rough comparison of the physical diagnostics over two crudely divided spatial scales, the innermost $10\arcsec \times 10\arcsec$ region versus the remaining outer, flanking region, to validate that the unique PAH characteristics of LLAGNs arise from physical processes relevant to the active nucleus.  Figure~\ref{fig:R_nuc_R_peri} shows that both diagnostics (sensitive to shocks and to the combined effects of shocks and radiative excitation) are statistically stronger toward the inner region of the galaxy, suggesting that they indeed appear to be associated intimately with the active nucleus.

Our discussion has focused on the role of AGNs in reducing the strength of PAH emission or in modifying its detailed relative band ratios. In principle, the AGN itself can also excite PAH molecules (\citealt{Jensen et al. 2017}), raising concern over the reliability of SFRs in AGNs inferred from PAH emission, especially as applied to individual features such as the 11.3\,$\mum$ band (e.g., \citealt{Diamond-Stanic & Rieke 2010, Alonso-Herrero et al. 2014, Esquej et al. 2014}). We note, however, that biases incurred from the selective suppression and/or enhancement of individual PAH features can be mitigated by using SFR calibrations based on the integrated PAH emission instead of the strength of any single band (e.g., \citealt{Xie & Ho 2019}).  Integrated PAH emission appears to yield unbiased SFRs even for AGNs powerful enough to qualify as quasars \citep{Xie et al. 2021}.

\subsection{Implications for AGN Feedback}\label{sec:section4.3}

Having shown that the environment around active galaxies leaves an imprint on the PAH spectrum, we next investigate whether we can establish a more direct link with the physical properties of the AGN.  No clear trends can be seen with either AGN bolometric luminosity or Eddington ratio\footnote{For ease of presentation, four sources with upper limits in $L_{\rm bol}$ are omitted from the plot, but we have verified that excluding them does not change our conclusions.}, defined as $L_{\rm bol}/L_{\rm Edd}$, with $L_{\rm Edd} = 1.26\times10^{38}\,(M_{\rm BH}/M_{\odot})$ (Figure~\ref{fig:PAH_Ratio_AGN_AGNproperties}). As with the majority of the nearby galaxy population \citep{Ho 2008}, the SINGS sample contains almost exclusively AGNs of extremely low luminosities ($L_{\rm bol} \approx 10^{39}-10^{42}$\,$\rm erg\ s^{-1}$) and Eddington ratios ($L_{\rm bol}/L_{\rm Edd} \approx 10^{-6} - 10^{-3}$). Such highly sub-Eddington systems, dominated by LINERs, have tiny mass accretion rates and extremely low radiative efficiency (\citealt{Ho et al. 2003, Ho 2009}).  As evidenced by their unique SEDs (i.e., \citealt{Ho 1999, Di Matteo et al. 2003, Pellegrini et al. 2003}) and in the inverse correlation between radio-loudness and Eddington ratio (\citealt{Ho 2002, Ho 2008, Terashima & Wilson 2003, Sikora et al. 2007}), LLAGNs redirect most of their accretion power from radiation to kinetic energy.  Interestingly, the AGN spaxels with the larger 11.3~$\mum$/7.7~$\mum$ ratios---those that we consider to show the most convincing evidence for shocks (Section~4.2)---coincide with LINERs that have among the lowest values of $L_{\rm bol}/L_{\rm Edd}$.  In short, we are witnessing the consequences of mechanical energy injection into the interstellar medium, presumably associated with jets and outflows launched by radiatively inefficient accretion flows \citep{Yuan & Narayan 2014}. Distinct from luminous, highly accreting supermassive BHs, whose radiative energy efficiently expels the surrounding gas through ``quasar-mode'' feedback (e.g., \citealt{Di Matteo et al. 2005, Hopkins et al. 2008}), low-level nuclear activity heats the gas through shocks (e.g., \citealt{McNamara & Nulsen 2007}). This form of ``kinetic-mode'' feedback has been widely embraced in recent cosmological simulations of galaxy evolution (\citealt{Weinberger et al. 2017, Dave et al. 2019}). Here, we suggest that kinetic-mode is directly responsible for the destruction of small PAH molecules.

It is illuminating to compare directly our sample of LLAGNs with powerful AGNs, where quasar-mode feedback might operate. The recent study of \cite{Xie & Ho 2022} finds that low-redshift quasars also exhibit distinct characteristics in their PAH spectrum, but their properties differ from those of the sources considered here.  As with LLAGNs, quasars also have smaller values of 6.2~$\mum$/7.7~$\mum$ compared to SFGs, but their differences are even more extreme (median value of 0.09 for quasars versus 0.25 for LLAGNs; see Figure~8 of Xie \& Ho 2022). Most notably, quasars in general have 11.3~$\mum$/7.7~$\mum$ ratios that are quite similar to those of SFGs, and lie close to the theoretical sequence of ionized PAHs with $N_{\rm C} \gtrsim 1000$, much larger than attained in LLAGNs.  Although the analysis of the quasars was based on globally integrated instead of spatially resolved spectra, quasars seem to lack the very large values of 11.3~$\mum$/7.7~$\mum$ reached by LLAGNs.  If, as we argue, the highly elevated values of 11.3~$\mum$/7.7~$\mum$ are indicative of shocks, we are tempted to conclude that shocks are less prevalent in quasars than in LLAGNs. While in both situations the PAH size distribution is depleted of smaller grains, the underlying causes differ, and the PAH size distribution of quasars shifts further toward larger grains. We argue that shocks directly destroy small grains in LLAGNs, especially the ionized grains, but the powerful radiation field in quasars deplete more small grains and keep the surviving larger grains ionized.

Lastly, apart from the above subtle variations imparted on the PAH spectrum, the two extremes of AGN activity also manifest themselves differently in the degree to which they impact the overall level of PAH emission.  Among LLAGNs, the PAH suppression becomes most severe at $L_{\rm FUV} \lesssim 10^{42}$\,$\rm erg\ s^{-1}$ (Figure~\ref{fig:PAH_FUV_Ne2}b), which corresponds to ${\rm SFR} \approx 0.05$\,$\rm M_\odot\ yr^{-1}$ \citep{Hao et al. 2011}, reaching values as extreme as $\Delta \log\, L_{\rm PAH} \approx -3$~dex. The magnitude of the PAH deficit increases with shock strength, albeit with large scatter (Figure~\ref{fig:PAH_deficit}a), and especially in response to the joint effect of shocks and radiation (Figure~\ref{fig:PAH_deficit}b). By stark contrast, quasars only begin to exhibit a PAH deficit among the most luminous sources ($L_{\rm bol} \gtrsim 10^{46}$\,$\rm erg\ s^{-1}$) with the highest levels of star formation activity (${\rm SFR} \gtrsim 50$\,$\rm M_\odot\ yr^{-1}$; see Figure~6 in \citealt{Xie & Ho 2022}).  And even then, the amount of PAH reduction is only $\Delta \log\, L_{\rm PAH} \lesssim -0.5$~dex, markely milder than in LLAGNs.

\section{Summary}\label{section:sec5}

We apply the methodology of \cite{Zhang et al. 2021} to perform a comprehensive study of the spatially resolved mid-IR Spitzer spectra of 66 nearby galaxies for which mapping-mode IRS observations are available from the SINGS survey.  Comprisingly roughly half in low-luminosity AGNs, the SINGS sample is ideal to investigate the effects of weak AGN activity on the properties of PAH emission. When combined with the recent results of \cite{Xie & Ho 2022}, we aim to use PAH emission as a diagnostic of AGN-related energy feedback on the interstellar medium, covering an unprecedented dynamic range in BH accretion rate, from the most feeble LINERs to powerful quasars. Combining the low-resolution spectra with near-IR and mid-IR images, we obtain robust SEDs that allow us to derive measurements of the total integrated ($5-20\,\mum$) PAH emission, as well as the emission from the individual 6.2, 7.7, and 11.3\,$\mum$ bands. Although more limited in spatial coverage, the high-resolution spectra furnish ionic fine-structure lines and molecular hydrogen lines that provide valuable physical diagnostics.

Our main conclusions are as follows:

\begin{enumerate}

\item In star-forming regions, the integrated PAH emission correlates tightly with the extinction-corrected FUV luminosity, with an intrinsic scatter of $\rm 0.24$ dex. Although the sub-linear correlation suggests that PAHs may be excited more than just young stars, the tight correlation confirms that PAH emission serves as an effective SFR indicator in star-forming environments.

\item Relative to the star-forming regions, the spatially resolved regions of low-luminosity AGNs, especially those with highly sub-Eddington luminosities, have a tendency to exhibit weaker PAH emission for a given FUV luminosity, as well as on average smaller 6.2\,$\mum$/7.7\,$\mum$ and larger 11.3\,$\mum$/7.7\,$\mum$ PAH band ratios.

\item Radiative effects alone cannot account for the observed trends. Instead, shocks, through the selective destruction of small grains, are plausibly the dominant agent responsible for the unique PAH spectra of low-luminosity AGNs, although radiative effects may also contribute a secondary role.

\item By comparison, quasars only suppress PAH emission by a modest level at very high luminosities, and their PAH band ratios reflect the influence of the strong radiation field in ionizing and destroying the grains.

\item The contrasting PAH characteristics of quasars and low-luminosity AGNs can serve as a potential diagnostic to differentiate the effects of quasar-mode versus kinetic-mode AGN feedback.

\end{enumerate}

\acknowledgments
We thank the anonymous referee for helpful comments and suggestions. This work was supported by the National Science Foundation of China (11721303, 11991052), China Manned Space Project (CMS-CSST-2021-A04, CMS-CSST-2021-A06), and the National Key R\&D Program of China (2016YFA0400702).


\begin{thebibliography}{}
\expandafter\ifx\csname natexlab\endcsname\relax\def\natexlab#1{#1}\fi

\bibitem[Aitken \& Roche(1985)]{Aitken & Roche 1985} Aitken, D.~K., \& Roche, P.~F.\ 1985, \mnras, 213, 777

\bibitem[Alatalo et al.(2015)]{Alatalo et al. 2015} Alatalo, K., Lacy, M., Lanz, L., et al.\ 2015, \apj, 798, 31

\bibitem[Allain et al.(1996a)]{Allain et al. 1996a} Allain, T., Leach, S., \& Sedlmayr, E.\ 1996a, \aap, 305, 602

\bibitem[Allain et al.(1996b)]{Allain et al. 1996b} Allain, T., Leach, S., \& Sedlmayr, E.\ 1996b, \aap, 305, 616

\bibitem[Allamandola et al.(1985)]{Allamandola et al. 1985} Allamandola, L.~J., Tielens, A.~G.~G.~M., \& Barker, J.~R.\ 1985, \apjl, 290, L25

\bibitem[Allamandola et al.(1989)]{Allamandola et al. 1989} Allamandola, L.~J., Tielens, A.~G.~G.~M., \& Barker, J.~R.\ 1989, \apjs, 71, 733


\bibitem[Alonso-Herrero et al.(2014)]{Alonso-Herrero et al. 2014} Alonso-Herrero, A., Ramos Almeida, C., Esquej, P., et al.\ 2014, \mnras, 443, 2766

\bibitem[Bakes \& Tielens(1994)]{Bakes & Tielens 1994} Bakes, E.~L.~O., \& Tielens, A.~G.~G.~M.\ 1994, \apj, 427, 822

\bibitem[Barth et al.(2009)]{Barth et al. 2009} Barth, A.~J., Strigari, L.~E., Bentz, M.~C., et al.\ 2009, \apj, 690, 1031

\bibitem[Bruzual \& Charlot(2003)]{Bruzual & Charlot 2003} Bruzual, G., \& Charlot, S.\ 2003, \mnras, 344, 1000

\bibitem[Calzetti et al.(2005)]{Calzetti et al. 2005} Calzetti, D., Kennicutt, R.~C., Bianchi, L., et al.\ 2005, \apj, 633, 871

\bibitem[Calzetti et al.(2007)]{Calzetti et al. 2007} Calzetti, D., Kennicutt, R.~C., Engelbracht, C.~W., et al.\ 2007, \apj, 666, 870

\bibitem[Cardelli et al.(1989)]{Cardelli et al. 1989} Cardelli, J.~A., Clayton, G.~C., \& Mathis, J.~S.\ 1989, \apj, 345, 245

\bibitem[Chabrier(2003)]{Chabrier 2003} Chabrier, G.\ 2003, \pasp, 115, 763

\bibitem[Chen et al.(2019)]{Chen et al. 2019} Chen, J., Shi, Y., Dempsey, R., et al. 2019, \mnras, 489, 855

\bibitem[Cisternas et al.(2013)]{Cisternas et al. 2013} Cisternas, M., Gadotti, D.~A., Knapen, J.~H., et al.\ 2013, \apj, 776, 50

\bibitem[Combes et al.(2019)]{Combes et al. 2019} Combes, F., Garc{\'\i}a-Burillo, S., Audibert, A., et al.\ 2019, \aap, 623, A79

\bibitem[Dale et al.(2017)]{Dale et al. 2017} Dale, D.~A., Cook, D.~O., Roussel, H., et al.\ 2017, \apj, 837, 90

\bibitem[Dale et al.(2007)]{Dale et al. 2007} Dale, D.~A., Gil de Paz, A., Gordon, K.~D., et al.\ 2007, \apj, 655, 863

\bibitem[Dale et al.(2006)]{Dale et al. 2006} Dale, D.~A., Smith, J.~D.~T., Armus, L., et al.\ 2006, \apj, 646, 161

\bibitem[Dale et al.(2009)]{Dale et al. 2009} Dale, D.~A., Smith, J.~D.~T., Schlawin, E.~A., et al.\ 2009, \apj, 693, 1821

\bibitem[Dav\'e et al.(2019)]{Dave et al. 2019} Dav\'e, R., Anglés-Alcázar, D., Narayanan, D., et al. 2019, \mnras, 486, 2827

\bibitem[Di Matteo et al.(2003)]{Di Matteo et al. 2003} Di Matteo, T., Allen, S.~W., Fabian, A.~C., et al.\ 2003, \apj, 582, 133

\bibitem[Di Matteo et al.(2005)]{Di Matteo et al. 2005} Di Matteo, T., Springel, V., \& Hernquist, L.\ 2005, \nat, 433, 604

\bibitem[Diamond-Stanic \& Rieke(2010)]{Diamond-Stanic & Rieke 2010} Diamond-Stanic, A.~M., \& Rieke, G.~H.\ 2010, \apj, 724, 140

\bibitem[Draine \& Li(2001)]{Draine & Li 2001} Draine, B.~T., \& Li, A.\ 2001, \apj, 551, 807

\bibitem[Draine \& Li(2007)]{Draine & Li 2007} Draine, B.~T., \& Li, A.\ 2007, \apj, 657, 810

\bibitem[Draine et al.(2021)]{Draine et al. 2021} Draine, B.~T., Li, A., Hensley, B.~S., et al.\ 2021, \apj, 917, 3

\bibitem[Ducci et al.(2014)]{Ducci et al. 2014} Ducci, L., Kavanagh, P.~J., Sasaki, M., et al.\ 2014, \aap, 566, A115

\bibitem[Dullo et al.(2020)]{Dullo et al. 2020} Dullo, B.~T., Bouquin, A.~Y.~K., Gil de Paz, A., et al.\ 2020, \apj, 898, 83

\bibitem[Engelbracht et al.(2007)]{Engelbracht et al. 2007} Engelbracht, C.~W., Blaylock, M., Su, K.~Y.~L., et al.\ 2007, \pasp, 119, 994

\bibitem[Esquej et al.(2014)]{Esquej et al. 2014} Esquej, P., Alonso-Herrero, A., Gonz{\'a}lez-Mart{\'\i}n, O., et al.\ 2014, \apj, 780, 86

\bibitem[Farrah et al.(2007)]{Farrah et al. 2007} Farrah, D., Bernard-Salas, J., Spoon, H.~W.~W., et al.\ 2007, \apj, 667, 149

\bibitem[Fazio et al.(2004)]{Fazio et al. 2004} Fazio, G.~G., Hora, J.~L., Allen, L.~E., et al.\ 2004, \apjs, 154, 10

\bibitem[F{\"o}rster Schreiber et al.(2004)]{Forster Schreiber et al. 2004} F{\"o}rster Schreiber, N.~M., Roussel, H., Sauvage, M., et al.\ 2004, \aap, 419, 501


\bibitem[Genzel et al.(1998)]{Genzel et al. 1998} Genzel, R., Lutz, D., Sturm, E., et al.\ 1998, \apj, 498, 579

\bibitem[Gil~de~Paz et al.(2007)]{Gil de Paz et al. 2007} Gil de Paz, A., Boissier, S., Madore, B.~F., et al.\ 2007, \apjs, 173, 185

\bibitem[Gliozzi et al.(2009)]{Gliozzi et al. 2009} Gliozzi, M., Satyapal, S., Eracleous, M., et al.\ 2009, \apj, 700, 1759

\bibitem[Gordon et al.(2008)]{Gordon et al. 2008} Gordon, K.~D., Engelbracht, C.~W., Rieke, G.~H., et al.\ 2008, \apj, 682, 336

\bibitem[Greene et al.(2020)]{Greene et al. 2020} Greene, J. E., Strader, J., \& Ho, L. C. 2020, ARA\&A, 58, 257

\bibitem[Guillard et al.(2012)]{Guillard et al. 2012} Guillard, P., Ogle, P.~M., Emonts, B.~H.~C., et al.\ 2012, \apj, 747, 95

\bibitem[Hao et al. (2011)]{Hao et al. 2011} Hao, C.-N., Kennicutt Jr., R. C., Johnson, B. D., et al. 2011, \apj, 741, 124

\bibitem[Heckman(1980)]{Heckman 1980} Heckman, T. M. 1980, A\&A, 87, 152

\bibitem[Hirashita et al.(2020)]{Hirashita et al. 2020} Hirashita, H., Deng, W., \& Murga, M.~S.\ 2020, \mnras, 499, 3046

\bibitem[Hirashita \& Murga(2020)]{Hirashita & Murga 2020} Hirashita, H., \& Murga, M.~S.\ 2020, \mnras, 492, 3779

\bibitem[Ho(1999)]{Ho 1999} Ho, L.~C.\ 1999, \apj, 516, 672

\bibitem[Ho(2002)]{Ho 2002} Ho, L.~C.\ 2002, \apj, 564, 120

\bibitem[Ho(2008)]{Ho 2008} Ho, L.~C.\ 2008, \araa, 46, 475

\bibitem[Ho(2009)]{Ho 2009} Ho, L.~C.\ 2009, \apj, 699, 626

\bibitem[Ho et al.(1997a)]{Ho et al. 1997a} Ho, L.~C., Filippenko, A.~V., \& Sargent, W.~L.~W.\ 1997a, \apjs, 112, 315

\bibitem[Ho et al.(2003)]{Ho et al. 2003} Ho, L.~C., Filippenko, A.~V., \& Sargent, W.~L.~W.\ 2003, \apj, 583, 159

\bibitem[Ho et al.(1997b)]{Ho et al. 1997b} Ho, L. C., Filippenko, A. V., Sargent, W. L. W., \& Peng, C. Y. 1997b, \apjs, 112, 391

\bibitem[Ho et al.(2009)]{Ho et al. 2009} Ho, L.~C., Greene, J.~E., Filippenko, A.~V., et al.\ 2009, \apjs, 183, 1

\bibitem[Holm et al.(2011)]{Holm et al. 2011} Holm, A.~I.~S., Johansson, H.~A.~B., Cederquist, H., et al.\ 2011, \jcp, 134, 044301

\bibitem[Hopkins et al.(2008)]{Hopkins et al. 2008} Hopkins, P. F., Hernquist, L., Cox, T. J., \& Kere\v{s}, D. 2008, \apjs, 175, 35

\bibitem[Houck et al.(2004)]{Houck et al. 2004} Houck, J.~R., Roellig, T.~L., van Cleve, J., et al.\ 2004, \apjs, 154, 18

\bibitem[Hunt et al.(2010)]{Hunt et al. 2010} Hunt, L.~K., Thuan, T.~X., Izotov, Y.~I., et al.\ 2010, \apj, 712, 164

\bibitem[IRSA(2022)]{IRSA 2022} IRSA, 2022, Galactic Dust Reddening and Extinction, IPAC, doi:10.26131/IRSA537

\bibitem[Jarrett et al.(2003)]{Jarrett et al. 2003} Jarrett, T.~H., Chester, T., Cutri, R., et al.\ 2003, \aj, 125, 525

\bibitem[Jarrett et al.(2020)]{Jarrett et al. 2020} Jarrett, T.~H., Chester, T., Cutri, R., et al. 2020, 2MASS Large Galaxy Atlas, IPAC, doi:10.26131/IRSA122

\bibitem[Jarrett et al.(2013)]{Jarrett et al. 2013} Jarrett, T.~H., Masci, F., Tsai, C.~W., et al.\ 2013, \aj, 145, 6

\bibitem[Jensen et al.(2017)]{Jensen et al. 2017} Jensen, J.~J., H{\"o}nig, S.~F., Rakshit, S., et al.\ 2017, \mnras, 470, 3071

\bibitem[Kaneda et al.(2005)]{Kaneda et al. 2005} Kaneda, H., Onaka, T., \& Sakon, I.\ 2005, \apjl, 632, L83

\bibitem[Kelly(2007)]{Kelly 2007} Kelly, B.~C.\ 2007, \apj, 665, 1489

\bibitem[Kennicutt et al.(2003)]{Kennicutt et al. 2003} Kennicutt Jr., R.~C., Armus, L., Bendo, G., et al.\ 2003, \pasp, 115, 928

\bibitem[Lebouteiller et al.(2011)]{Lebouteiller et al. 2011} Lebouteiller, V., Bernard-Salas, J., Whelan, D.~G., et al.\ 2011, \apj, 728, 45

\bibitem[Leger \& Puget(1984)]{Leger & Puget 1984} Leger, A., \& Puget, J.~L.\ 1984, \aap, 500, 279

\bibitem[Lehmer et al.(2019)]{Lehmer et al. 2019} Lehmer, B.~D., Eufrasio, R.~T., Tzanavaris, P., et al.\ 2019, \apjs, 243, 3

\bibitem[Li(2020)]{Li 2020} Li, A.\ 2020, Nature Astronomy, 4, 339

\bibitem[Li \& Draine(2002)]{Li & Draine 2002} Li, A. \& Draine, B.~T.\ 2002, \apj, 576, 762

\bibitem[Liu et al. (2011)]{Liu et al. 2011}Liu, G., Koda, J., Calzetti, D., et al. 2011, ApJ, 735, 63

\bibitem[Maoz et al.(1995)]{Maoz et al. 1995}  Maoz, D., Filippenko, A. V., Ho, L. C., et al. 1995, \apj, 440, 91

\bibitem[Maragkoudakis et al.(2018)]{Maragkoudakis et al. 2018} Maragkoudakis, A., Ivkovich, N., Peeters, E., et al.\ 2018, \mnras, 481, 5370

\bibitem[McNamara \& Nulsen (2007)]{McNamara & Nulsen 2007} McNamara, B. R., \& Nulsen, P. E. J. 2007, ARA\&A, 45, 117

\bibitem[Micelotta et al.(2010a)]{Micelotta et al. 2010a} Micelotta, E.~R., Jones, A.~P., \& Tielens, A.~G.~G.~M.\ 2010a, \aap, 510, A36

\bibitem[Micelotta et al.(2010b)]{Micelotta et al. 2010b} Micelotta, E.~R., Jones, A.~P., \& Tielens, A.~G.~G.~M.\ 2010b, \aap, 510, A37

\bibitem[Molina et al.(2022)]{Molina et al. 2022} Molina, J., Ho, L. C., Wang, R., et al. 2022, \apj, submitted

\bibitem[Morrissey et al.(2007)]{Morrissey et al. 2007} Morrissey, P., Conrow, T., Barlow, T.~A., et al.\ 2007, \apjs, 173, 682

\bibitem[Moustakas et al.(2010)]{Moustakas et al. 2010} Moustakas, J., Kennicutt, R.~C., Tremonti, C.~A., et al.\ 2010, \apjs, 190, 233

\bibitem[Murga et al.(2016)]{Murga et al. 2016} Murga, M.~S., Khoperskov, S.~A., \& Wiebe, D.~S.\ 2016, Astronomy Reports, 60, 233

\bibitem[Murga et al.(2019)]{Murga et al. 2019} Murga, M.~S., Wiebe, D.~S., Sivkova, E.~E., et al.\ 2019, \mnras, 488, 965

\bibitem[NASA/IPAC Infrared Science Archive(2020)]{IPAC 2020} NASA/IPAC Infrared Science Archive, 2020, WISE All-Sky 4-band Atlas Coadded Images, IPAC, doi:10.26131/IRSA151

\bibitem[O'Dowd et al.(2009)]{ODowd et al. 2009} O'Dowd, M.~J., Schiminovich, D., Johnson, B.~D., et al.\ 2009, \apj, 705, 885

\bibitem[O'Halloran et al.(2006)]{OHalloran et al. 2006} O'Halloran, B., Satyapal, S., \& Dudik, R. P. 2006, \apj, 641, 795

\bibitem[Peeters et al.(2004)]{Peeters et al. 2004} Peeters, E., Spoon, H.~W.~W., \& Tielens, A.~G.~G.~M.\ 2004, \apj, 613, 986

\bibitem[Pellegrini(2010)]{Pellegrini 2010} Pellegrini, S.\ 2010, \apj, 717, 640

\bibitem[Pellegrini et al.(2003)]{Pellegrini et al. 2003} Pellegrini, S., Venturi, T., Comastri, A., et al.\ 2003, \apj, 585, 677

\bibitem[Pereira-Santaella et al.(2017)]{Pereira-Santaella et al. 2017} Pereira-Santaella, M., Rigopoulou, D., Farrah, D., et al.\ 2017, \mnras, 470, 1218

\bibitem[Rigopoulou et al.(2021)]{Rigopoulou et al. 2021} Rigopoulou, D., Barale, M., Clary, D., et al.\ 2021, \mnras, 504, 5287

\bibitem[Roussel et al.(2007)]{Roussel et al. 2007} Roussel, H., Helou, G., Hollenbach, D.~J., et al.\ 2007, \apj, 669, 959

\bibitem[Roussel et al.(2006)]{Roussel et al. 2006} Roussel, H., Helou, G., Smith, J.~D., et al.\ 2006, \apj, 646, 841

\bibitem[Sales et al.(2010)]{Sales et al. 2010} Sales, D.~A., Pastoriza, M.~G., \& Riffel, R.\ 2010, \apj, 725, 605

\bibitem[Schlafly \& Finkbeiner(2011)]{Schlafly & Finkbeiner 2011} Schlafly, E.~F., \& Finkbeiner, D.~P.\ 2011, \apj, 737, 103

\bibitem[Shangguan et al.(2018)]{Shangguan et al. 2018} Shangguan, J., Ho, L.~C., \& Xie, Y.\ 2018, \apj, 854, 158

\bibitem[Shipley et al.(2016)]{Shipley et al. 2016} Shipley, H.~V., Papovich, C., Rieke, G.~H., et al.\ 2016, \apj, 818, 60

\bibitem[SINGS Team(2020)]{SINGS 2020} SINGS Team, 2020, Spitzer Infrared Nearby Galaxy Survey, IPAC, doi:10.26131/IRSA424

\bibitem[Sikora et al.(2007)]{Sikora et al. 2007} Sikora, M., Stawarz, L., \& Lasota, J.-P. 2007, \apj, 658, 815

\bibitem[Skrutskie et al.(2006)]{Skrutskie et al. 2006} Skrutskie, M.~F., Cutri, R.~M., Stiening, R., et al.\ 2006, \aj, 131, 1163

\bibitem[Smith et al.(2007a)]{Smith et al. 2007a} Smith, J.~D.~T., Armus, L., Dale, D.~A., et al.\ 2007a, \pasp, 119, 1133

\bibitem[Smith et al.(2007b)]{Smith et al. 2007b} Smith, J.~D.~T., Draine, B.~T., Dale, D.~A., et al.\ 2007b, \apj, 656, 770

\bibitem[Stierwalt et al.(2014)]{Stierwalt et al. 2014} Stierwalt, S., Armus, L., Charmandaris, V., et al.\ 2014, \apj, 790, 124

\bibitem[Taniguchi et al.(1996)]{Taniguchi et al. 1996} Taniguchi, Y., Ohyama, Y., Yamada, T., Mouri, H., \& Yoshida, M. 1996, \apj, 467, 215

\bibitem[Terashima \& Wilson (2003)]{Terashima & Wilson 2003} Terashima, Y., \& Wilson, A. S. 2003, \apj, 583, 145

\bibitem[Thater et al.(2019)]{Thater et al. 2019} Thater, S., Krajnovi{\'c}, D., Cappellari, M., et al.\ 2019, \aap, 625, A62

\bibitem[Thornley et al.(2000)]{Thornley et al. 2000} Thornley, M.~D., F{\"o}rster Schreiber, N.~M., Lutz, D., et al.\ 2000, \apj, 539, 641

\bibitem[Tielens(2005)]{Tielens 2005} Tielens, A.~G.~G.~M.\ 2005, The Physics and Chemistry of the Interstellar Medium (Cambridge, UK: Cambridge Univ. Press)

\bibitem[Tielens(2008)]{Tielens 2008} Tielens, A.~G.~G.~M.\ 2008, \araa, 46, 289

\bibitem[Treyer et al.(2010)]{Treyer et al. 2010} Treyer, M., Schiminovich, D., Johnson, B.~D., et al.\ 2010, \apj, 719, 1191

\bibitem[Turner et al.(1977)]{Turner et al. 1977} Turner, J., Kirby-Docken, K., \& Dalgarno, A.\ 1977, \apjs, 35, 281

\bibitem[Vega et al.(2010)]{Vega et al. 2010} Vega, O., Bressan, A., Panuzzo, P., et al.\ 2010, \apj, 721, 1090

\bibitem[Voit(1992)]{Voit 1992} Voit, G.~M.\ 1992, \mnras, 258, 841

\bibitem[Weinberger et al.(2017)]{Weinberger et al. 2017} Weinberger, R., Springel, V., Hernquist, L., et al.\ 2017, \mnras, 465, 3291

\bibitem[Weingartner \& Draine(2001b)]{Weingartner & Draine 2001b} Weingartner, J.~C., \& Draine, B.~T.\ 2001b, \apjs, 134, 263

\bibitem[Weingartner \& Draine(2001a)]{Weingartner & Draine 2001a} Weingartner, J.~C., \& Draine, B.~T.\ 2001a, \apj, 548, 296

\bibitem[Werner et al.(2004)]{Werner et al. 2004} Werner, M.~W., Roellig, T.~L., Low, F.~J., et al.\ 2004, \apjs, 154, 1

\bibitem[Wright et al.(2010)]{Wright et al. 2010} Wright, E.~L., Eisenhardt, P.~R.~M., Mainzer, A.~K., et al.\ 2010, \aj, 140, 1868

\bibitem[Wu et al.(2005)]{Wu et al. 2005} Wu, H., Cao, C., Hao, C.-N., et al.\ 2005, \apjl, 632, L79

\bibitem[Xie \& Ho(2019)]{Xie & Ho 2019} Xie, Y., \& Ho, L.~C.\ 2019, \apj, 884, 136

\bibitem[Xie \& Ho(2022)]{Xie & Ho 2022} Xie, Y., \& Ho, L.~C.\ 2022, \apj, 925, 218

\bibitem[Xie et al.(2018)]{Xie et al. 2018} Xie, Y., Ho, L.~C., Li, A., et al.\ 2018, \apj, 860, 154

\bibitem[Xie et al.(2021)]{Xie et al. 2021} Xie, Y., Ho, L. C., Zhuang, M.-Y., \& Shangguan, J. 2021, \apj, 910, 124

\bibitem[Yuan \& Narayan (2014)]{Yuan & Narayan 2014} Yuan, F., \& Narayan, R. 2014, ARA\&A, 52, 529

\bibitem[Zhang \& Ho(2022)]{Zhang & Ho 2022} Zhang, L., \& Ho, L.~C.\ 2022, \apj, in press

\bibitem[Zhang et al.(2021)]{Zhang et al. 2021} Zhang, L., Ho, L.~C, \& Xie, Y.\ 2021, \aj, 161, 29

\end{thebibliography}
\end{document}